\pdfoutput=1
\documentclass[aps,prx,superscriptaddress,%
twocolumn%
]{revtex4-2}
\usepackage[usenames,dvipsnames]{xcolor}

\usepackage[normalem]{ulem}

\usepackage{graphicx}
\usepackage{amsmath}
\usepackage{amssymb}
\usepackage{float}
\usepackage{bm}
\usepackage{wasysym}
\usepackage{longtable}
\usepackage{multirow}
\usepackage{colortbl}
\usepackage{pifont} 
\newcommand{\cmark}{\color{ForestGreen}\ding{51}}%
\newcommand{\xmark}{\color{red}\ding{55}}%
\usepackage[colorlinks=true,linkcolor=blue,citecolor=blue,urlcolor=blue]{hyperref}

\AtBeginDocument{\usepackage{booktabs}}

\begin{document}

\title{Shear and breathing modes of layered materials}

\keywords{layered materials, Raman, infrared, multilayer, fan diagrams}
\author{Giovanni Pizzi}
\email{giovanni.pizzi@epfl.ch}
\affiliation{Theory and Simulation of Materials (THEOS), and National Centre for Computational Design and Discovery of Novel Materials (MARVEL), \'Ecole Polytechnique F\'ed\'erale de Lausanne, CH-1015 Lausanne, Switzerland}
\author{Silvia Milana}
\affiliation{Cambridge Graphene Centre, University of Cambridge, Cambridge CB3 OFA, UK}
\author{Andrea C. Ferrari}
\affiliation{Cambridge Graphene Centre, University of Cambridge, Cambridge CB3 OFA, UK}
\author{Nicola Marzari}
\affiliation{Theory and Simulation of Materials (THEOS), and National Centre for Computational Design and Discovery of Novel Materials (MARVEL), \'Ecole Polytechnique F\'ed\'erale de Lausanne, CH-1015 Lausanne, Switzerland}
\author{Marco Gibertini}
\email{marco.gibertini@unimore.it}
\affiliation{Theory and Simulation of Materials (THEOS), and National Centre for Computational Design and Discovery of Novel Materials (MARVEL), \'Ecole Polytechnique F\'ed\'erale de Lausanne, CH-1015 Lausanne, Switzerland}
\affiliation{Department of Quantum Matter Physics, University of Geneva, CH-1211 Gen\`eve, Switzerland}
\affiliation{Dipartimento di Scienze Fisiche, Informatiche e Matematiche, University of Modena and Reggio Emilia, IT-41125 Modena, Italy}

\begin{abstract}
Layered materials (LMs), such as graphite, hexagonal boron nitride, and transition-metal dichalcogenides, are at the centre of an ever increasing research effort, due to their scientific and technological relevance. Raman and infrared spectroscopies are accurate, non-destructive, approaches to determine a wide range of properties, including the number of layers and the strength of the interlayer interactions. Here, we present a general approach to predict the complete spectroscopic fan diagrams, i.e., the relations between frequencies and number of layers, $N$, for the optically-active shear and layer-breathing modes of any multilayer comprising $N\geq2$ identical layers. In order to achieve this, we combine a description of the normal modes in terms of a one-dimensional mechanical model, with symmetry arguments that describe the evolution of the point group as a function of $N$. Group theory is then used to identify which modes are Raman and/or infrared active, and to provide diagrams of the optically-active modes for any stack composed of identical layers. We implement the method and algorithms in an open-source tool directly available on the Materials Cloud portal, to assist any researcher in the prediction and interpretation of such diagrams. Our work will underpin all future efforts on Raman and Infrared characterization of known, and yet not investigated, LMs.
\end{abstract}
\maketitle
\section{Introduction}
Layered materials (LMs) are at the centre of an ever growing research effort due to the variety of their potential applications in a wide range of fields\cite{Ferrari2015}. There are at least 5000 materials that are layered\cite{Mounet2018} with at least 1800 that are readily exfoliable\cite{Mounet2018,Cheon2017,Ashton2017,Choudhary2017}, and even more that could be synthesised\cite{Haastrup2018,Zhou2019,Backes_2020,Taghizadeh2020}.
However, only a very small fraction of these have been experimentally investigated to date, such as graphene, hexagonal boron nitride (hBN), black phosphorous (BP), transition metal dichalcogenides (TMDs), InSe and other monochalcogenides, MAXenes, and very few others.
When a given LM is exfoliated into a multilayer (ML), the optical and electronic properties change with the number of layers ($N$).
For a given $N$, the properties can be tuned by varying the relative orientation of the layers\cite{Bistritzer2011,Cao2018a,Cao2018b,Wu2019}.
For a given $N$ and orientation, properties can also be changed by arranging different LMs together in heterostructures (LMHs)\cite{Heo2015,Nayak2017,Alexeev2019,Tran2019,Seyler2019,Jin2019}.
The degrees of freedom are such that it will take decades, if ever, before all possible LMs will be exfoliated and investigated when arranged in LMHs, as a function of $N$ and of relative orientation.
Due to the extraordinary range of properties that can be addressed, it is essential to develop approaches to readily identify $N$ in any given assembly or device.

Techniques to measure $N$ in a given sample based on optical contrast\cite{Casiraghi2007} have been developed.
However, they depend on the substrate and do not readily provide information such as strain or doping.
A more informative approach is offered by Raman\cite{Ferrari2013} and infrared (IR)\cite{Jiang2008} spectroscopies that probe phonons in a given material.

In particular, in LMs there are two fundamentally different sets of modes.
Those coming from the relative motion of the constituent atoms within each layer, usually found at high frequencies ($>$100~cm$^{-1}$)\cite{Ferrari2013}, and those due to relative motions of the layers themselves, either perpendicular, C modes, or parallel, layer breathing (LB) modes, to their normal\cite{Ferrari2013,Ji_review_2016,Zhang_review_2016,Liang_review_2017}.
Several studies have identified these modes in a limited set of ML-LMs, like ML-graphene\cite{Tan2012,Lui2012,Wu2014,Wu2015}, TMDs (such as MoS$_2$\cite{Zha2013,Boukhicha2013}, MoSe$_2$\cite{Kim2016}, WS$_2$\cite{Yang2017}, WSe$_2$\cite{Zhao2013}, MoTe$_2$\cite{Song2016,Fro2015}, ReS$_2$\cite{Nag2015,Zhao2015,Qiao2016}, PtS$_2$\cite{Zhao2016}), NbSe$_2$\cite{Xi2015,He2016,Orchin2019}, hBN\cite{Stenger2017}, phosphorene\cite{Ling2015,Luo2015bp,Dong2016}, Bi$_2$X$_3$\cite{Zha2014}, and  metal chalcogenides (like GaSe\cite{Lim2020,Molas2020}, InSe\cite{Molas2020}, and SnS$_2$\cite{Sriv2018}).

The optically-active (Raman or IR) modes can be plotted as a function of $N$, in a graph that represents a fan, and it is thus called a fan diagram\cite{Tan2012}. The experimental data can be explained with a linear chain model\cite{Tan2012}, whereby each plane is linked to the next one by a spring, modelled by scalar interlayer force constants corresponding to a motion parallel (C) or perpendicular (LB) to the planes\cite{Tan2012}.

Here, we extend the linear chain model to every possible exfoliable LM composed of identical layers by implementing a group-theory approach.
This starts from the bulk LM symmetry properties to derive a general tensorial expression for the interlayer force constants.
We show how to derive the evolution of the point group for any $N$ of a ML, knowing the space group of the bulk structure, considered as the repetition of a single layer stacked recursively.
This is then used to assign each normal mode to a given irreducible representation of the corresponding point group, in order to assess its optical activity and obtain the fan diagram of each LM.
Finally, we provide an online tool, available on the Materials Cloud\cite{Talirz2020} at the address \url{https://materialscloud.org/work/tools/layer-raman-ir}, that accepts user-supplied structures and computes on the fly the corresponding fan diagram and symmetry-compliant form of the interlayer force constants.
Our work provides the interpretation of the patterns that are measured in experimental fan diagrams of any LM composed of identical layers, either already experimentally investigated, or, more importantly, any of the thousands that will be studied in the future.
\section{Fan diagrams: prediction and interpretation}
\label{sec:fan-diagrams-theory}
A fan diagram is a plot of the normal-mode frequencies associated with the rigid relative motion of the layers in a ML-LM, as a function of $N$.
The fan diagram frequencies are a fingerprint of each material.
Their trend as a function of $N$ depends on the atomic structure and on the symmetry both of ML-LM system and of the corresponding bulk LM.

We now develop a theoretical model to interpret the experimental results and assess the origin and character of these vibrational modes and their expected optical activity.
Such a model needs a number of components.
\begin{enumerate}
\item We need an approach to compute the normal vibrational modes of ML-LMs and their frequencies, using a model that can capture the system geometry, and only depends on a few material parameters, such as the elastic constants between each pair of layers.

\item We need to numerically identify and extract the layers of a ML-LMs from the bulk structure and analyse their crystal symmetry.
Given the space group of bulk LMs, it is possible to determine all possible symmetries of ML-LMs system with a given $N$.

\item We need to exploit the symmetry information to identify the optical activity of each normal mode (i.e., if the mode is Raman or IR active, and, if so, if it can be detected in the most commonly used back-scattering geometry\cite{Ferrari2006,Ferrari2013}).
We achieve this by using group theory to classify each mode, assigning it to the irreducible representation to which it belongs, thus determining its optical activity.

\item We then combine points 1-3 above in a single model to enable the interpretation of the experimental data.
\end{enumerate}

This paper is organised such that a reader interested only in the applications of this method can skip the rest of this Sec.~\ref{sec:fan-diagrams-theory} and move directly to the results reported in Sec.~\ref{sec:results} and to the description of the free online tool, presented in Sec.~\ref{sec:tool}, that applies this approach to any user-provided LM.

\subsection{Definition of a layered material and nomenclature}
\label{sec:definitions}
We are interested in modelling the vibrational properties of LMs when layers move essentially as rigid units as a consequence of the strong covalent bonds between atoms in a given layer, as opposed to the weak van-der-Waals interactions keeping layers together.

In this limit, vibrations can be described in terms of interlayer force constants, acting as restoring forces between nearby layers.

In order to limit the number of parameters in the model and make use of crystal symmetry and space-group concepts to seamlessly predict the normal modes and their optical activity, we consider LMs with a sufficiently regular stacking (to be described below, in particular focusing on LMs composed of identical layers), which covers the majority of naturally occurring LMs.

In this paper we cover any multilayer comprising $N\geq2$ identical layers. We note that our approach can be numerically extended to any LM and LMH. Stacking in LMHs lowers the symmetry, lifting most symmetry constraints on the optical activity of modes. Group theory alone could predict modes to be active even if the corresponding intensity might be negligible; thus further computation of the optical-coupling matrix elements becomes essential.
In non-recursive stacking sequences, more parameters enter the description of interlayer force constants (with a different force-constant matrix for each layer pair and for each possible relative orientation of the two), which can be extracted from additional first-principles simulations.

We follow a practical approach, giving a brief and intuitive explanation of the important symmetry properties of these LMs.
Ref.~\onlinecite{Durovic2006} provides a complete treatment with formal definitions and proofs.
Since the nomenclature used in the experimental literature of ML fan diagrams often differs from that used in the crystallographic community\cite{Durovic2006}, we also provide a mapping between the names used in the two communities, where appropriate.

ML-LMs are called polytypes by the International Union of Crystallography (see Ref.~\onlinecite{Guinier1984} for a formal definition).
A theory to describe these ML-LMs, based solely on the symmetry of each layer and on the symmetry relation between subsequent layers, was developed in Refs.~\onlinecite{Dornberger1964,Dornberger1972}.

We limit our study to LMs where all layers are identical and can be mapped onto each other through coincidence operations, defined as isometries (i.e., space transformations that preserve the distance between any two points) bringing a layer of the ML-LM onto the next one.
As already noted above this excludes, e.g., bulk LMHs formed by different LMs, as in the case of franckeite\cite{Molina-Mendoza2017,Velicky2017} but is the typical case for exfoliable materials.

\begin{figure}
\centerline{\includegraphics[width=90mm]{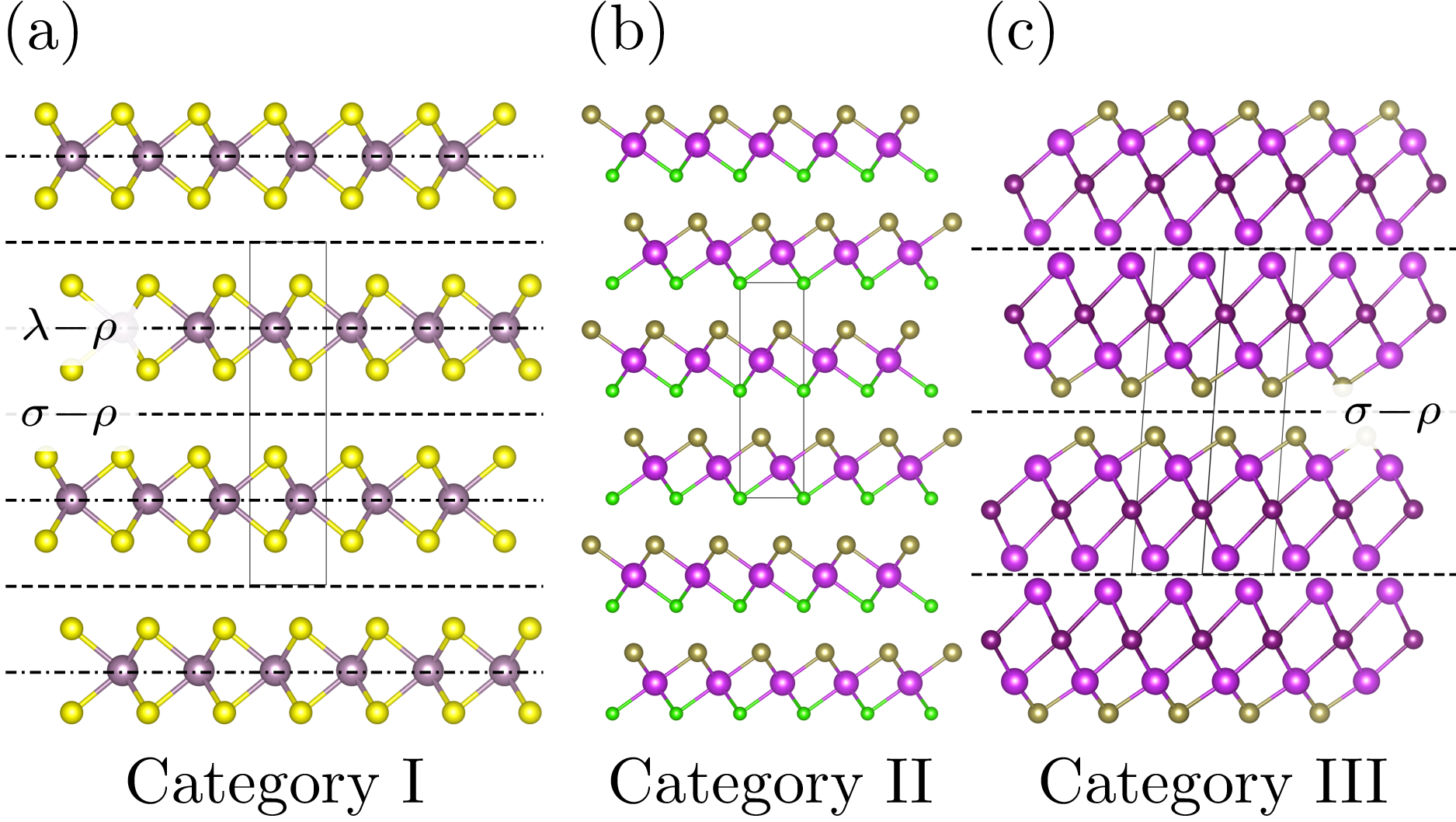}}
\caption{\label{fig:MDO-categories} 3 categories of LMs allowed from a symmetry point of view, for MDO polytypes of equivalent layers.
(a) Category I: each layer is non-polar along the stacking direction (i.e., it has a symmetry operation that flips it upside down), such as in MoS$_2$ (structure from the Crystallography Open Database (COD\cite{COD}), code 9007660).
Mo atoms are shown in violet and S atoms in yellow.
(b) Category II: each layer is polar along the stacking direction, and all layers are oriented in the same direction, such as in BiTeCl (structure from the Inorganic Crystal Structure Database (ICSD\cite{ICSD}), code 79362).
Bi atoms are shown in light purple, Te atoms in brown, and Cl atoms in green.
(c) Category III: each layer is polar along the stacking direction, and they stack in alternating polarisation directions, such as Bi$_2$TeI (ICSD\cite{ICSD} code 153858).
Bi atoms are shown in light purple, Te atoms in brown, and I atoms in dark purple.
Symmetry planes for layer-order-changing operations ($\rho$ planes) are indicated with dashed lines ($\sigma$ operations) or dotted-dashed lines ($\lambda$ operations).
Note that the LM in Category III is for illustrative purposes: depending on the nature chemical bonding, this could be considered as 3 non-equivalent layers (2 BiTeI layers analogous to those of Category II, and one layer of Bi atoms)}
\end{figure}

The coincidence operation that brings one layer onto the next might not be the same for all layers, e.g., if the first layer is mapped onto the second one by a translation, while the second is mapped onto the third by a  rotation.
Again with the goal to limit the number of parameters in the model, we then consider an additional requirement by limiting our analysis to MDO (maximum degree of order) polytypes.
These are LMs where the coincidence operation is total, i.e., it is the same between any pair of adjacent layers.
As a consequence\cite{Durovic2006}, in a MDO polytype any triplet of subsequent layers is equivalent, while it is not true that every pair is equivalent, as we shown in the example of Fig.~\ref{fig:MDO-categories}c for Bi$_2$TeI.
As any triplet is equivalent, MDO polytypes have only 1 or 2 independent interatomic force-constant tensors that occur between nearby layers in the triplet, while all other tensors can be reconstructed using symmetry arguments.
If the coincidence operation is not total, the relative arrangement of atoms in pairs of subsequent layers could be very different, leading to different interactions between them, even if this almost never occurs in naturally occurring exfoliable materials.
\begin{table*}[tbp]
\caption{\label{tab:operations-reference-per-category}Summary table of the type of operations in each of the categories of Fig.~\ref{fig:MDO-categories}.}
\begin{tabular}{cccccc}
    & & \multicolumn{2}{c}{$\lambda$ (operations of the monolayer)} & \multicolumn{2}{c}{$\sigma$ (operations bringing a layer onto the next one)} \\
    Category &  Polar layers & $\lambda-\rho$ ($\lambda$ LOC) & $\lambda-\tau$ ($\lambda$ non-LOC) & $\sigma-\rho$ ($\sigma$ LOC) & $\sigma-\tau$ ($\sigma$ non-LOC) \\
\hline\hline
\multirow{2}{*}{I}   & \multirow{2}{*}{\xmark} & \cmark & \multirow{2}{*}{\cmark} & \cmark & \multirow{2}{*}{\cmark} \\
   &  & (on layer plane) &  & (on middle plane) &  \\ \hline
II  & \cmark & \xmark & \cmark & \xmark & \cmark \\ \hline
\multirow{2}{*}{III}   & \multirow{2}{*}{\cmark} & \multirow{2}{*}{\xmark} & \multirow{2}{*}{\cmark} & \cmark & \multirow{2}{*}{\cmark} \\
& & & & (on middle plane) &
\end{tabular}

\end{table*}

Within these constraints, we can classify all LMs in 3 categories\cite{Durovic2006}, shown with 3 examples in Fig.~\ref{fig:MDO-categories}.

We first consider the case where each layer is non-polar along the stacking direction (i.e., it has a symmetry that flips it upside down) and then when it is polar.
In the non-polar case, only one possibility exists (Category I, Fig.~\ref{fig:MDO-categories}a).
In the polar case there are 2 options: either the polarity has the same orientation for all layers (Category II, Fig.~\ref{fig:MDO-categories}b) or it alternates between layers (Category III, Fig.~\ref{fig:MDO-categories}c).
These three categories have very different sets of symmetry operations.
We note that Category III, while considered here for completeness from a symmetry point of view, never occurs to the best of our knowledge for the most common LMs, such as graphite, hBN or the TMDs.

We define the planes of the layers in the LM as the ``horizontal'' direction, and the stacking direction of the LM as the ``vertical'' or $z$ one (note that the third vector of the bulk unit cell might not be orthogonal to the plane of the layers, see, e.g., Fig.~\ref{fig:MDO-categories}c).

We then distinguish the symmetry operations of the 1L-LM (called $\lambda$ symmetries in Ref.~\onlinecite{Durovic2006}) and the coincidence operations bringing a layer onto the next one ($\sigma$ symmetries).

Any symmetry operation can either change the sign of any vertical coordinate, i.e., flip the layer upside down (called $\rho$ operations\cite{Durovic2006}, like inversion,  roto-reflections, reflections under horizontal planes, or two-fold rotations with an horizontal axis) or not change the sign of vertical-direction coordinates (called $\tau$ operations\cite{Durovic2006}, e.g., translations, rotations with a vertical axis, or reflections under vertical planes; these form a subgroup).
Since, in a LM, all $\rho$ operations change the stacking order of the layers (e.g., a stacking 1-2-3-1-2-3 becomes 3-2-1-3-2-1), in the following we call them layer-order-changing (LOC) operations\cite{Scheuschner2015}, while we call the $\tau$ ones non-LOC operations.

With these definitions, for non-polar layers (Fig.~\ref{fig:MDO-categories}a) both $\lambda$ and $\sigma$ operations can be either LOC or non-LOC\cite{Durovic2006}.
Thus, we can formally define the vertical $z$ coordinate of each layer as the $z$ coordinate of its inversion centre (or reflection plane or rotation axis).
The plane with this $z$ coordinate is called the layer plane\cite{Durovic2006}.
Then, LOC operations can either be $\lambda$, and in this case their symmetry elements are on a layer plane, or $\sigma$, and they must lie on planes halfway between layer planes, as shown in Fig.~\ref{fig:MDO-categories}a.
We will call these planes ``middle planes''.
Non-LOC $\sigma$ operations (bringing one layer onto the next) are always combined with a translation along the vertical direction.

Polar layers do not have any symmetry operation that flips the $z$ coordinates, so all $\lambda$ operations are non-LOC.
However, while in Category II of Fig.~\ref{fig:MDO-categories} also all $\sigma$ coincidence operations are non-LOC (because the polarity direction is never reversed), in Category III all $\sigma$ coincidence operations must be LOC (changing thus polarisation orientation between consecutive layers) and they also lie on middle planes.
In Category III we cannot univocally define a layer plane, since there is no layer inversion plane, but by symmetry it is instead possible to define two sets of middle planes.

This distinction into 3 Categories is very important for the modelling of the interlayer force constants.
In Categories I, II, all pairs of layers are equivalent, therefore the same interlayer force constant matrix can be used to describe the interaction between any pair of layers.
In Category III there are 2 different interlayer force constant matrices, depending on whether the polarisations of the two neighbouring layers are pointing inwards or outwards with respect to the van-der-Waals gap between them (see more detailed discussions in Sec.~\ref{sec:normal-modes} and Appendix~\ref{app:grouping-catIII}).

When a LM satisfies all the conditions above the description of its vibrational properties and of the symmetries of ML-LM is greatly simplified and can be carried out analytically or semi-analytically.
For this reason in the rest of the paper we focus only on this class of LMs.
Nonetheless, our approach can be seamlessly extended to any form of LMs or even LMHs, although a numerical treatment might in general be needed, with a decreased predictivity owing to the increased number of free parameters.

\subsection{Obtaining the point group of a ML-LM}
\label{sec:multilayer-pointgroup}
We now consider how to obtain the point group of a ML-LM with $N$ layers (needed to predict its optical activity) given the point group of its parent bulk LM (B-LM), i.e., extended periodically in the direction orthogonal to the layers.

We call $n_c$ the number of layers in the B-LM conventional cell, and $n_p$ the number of layers in the B-LM primitive cell.
These two numbers are in general different, e.g., for centred cells or for rhombohedral systems.
For instance, rhombohedral graphite has $n_c=3$ and $n_p=1$.
We also define the ``stacking index'' of a layer as an integer indexing the layers so that, e.g., if a layer has stacking index $\ell$, then the next layer (in the positive vertical direction) has stacking index $\ell+1$.

The stacking direction, orthogonal to the planes of the layers, is unique.
Thus for some crystal systems it is prescribed by symmetry.
In particular, in tetragonal, hexagonal and trigonal systems, the stacking direction must be along the $n-$fold characteristic symmetry axis (e.g., the $c$ axis for tetragonal systems): if this was not the case, the $n-$fold rotations (with $n>2$) would bring the stacking direction into other distinct staking directions, which violates its unicity.
The same arguments imply that cubic systems are not compatible with a layered structure\cite{OKeeffe,Scheuschner2015}: if a given direction, say $z$, were the stacking direction, then also $x$ and $y$ should be by symmetry, as in cubic systems all principal directions are equivalent.
Therefore, we do not consider cubic systems in the rest of this work.

In orthorhombic, monoclinic and triclinic systems the stacking direction is instead not prescribed by symmetry, therefore the space group alone is not sufficient to characterise the system.
Instead, we need to consider all inequivalent settings, i.e., possible non-conventional choices for the origin and lattice vectors with respect to symmetry elements.
We consider all settings that are typically discussed in crystallography\cite{Souvignier_2016,Shmueli_2010},  identified by their Hall number\cite{Hall1981}.
This ranges from 1 to 488 if we exclude cubic systems.
E.g., space group 17 (P222$_1$, a primitive orthorhombic system with one screw axis and no mirror symmetry) can be realised in three different settings, depending on the direction of the screw axis, with Hall number 109, 110, and 111 for the screw axis aligned along the third, first and second cell axis, respectively.
In this work we are always assuming the stacking direction to be orthogonal to the first two lattice vectors.
Then, for setting P222$_1$ with Hall number 109, the $2_1$ screw axis is along the stacking direction, and a ML in this Hall setting has different symmetry properties than Hall numbers 110 and 111, corresponding to P2$_1$22 and P22$_1$2, that are equivalent for our purposes because in both cases the screw axis is horizontal.

We now present a strategy to obtain ``compatibility relations'', i.e., rules determining the possible point groups $G_N$ of a ML-LM as a function of $N$, by knowing the B-LM space group and setting (thus also the B-LM point group $G_b$), and the direction along which the material is layered.
This allows us to identify which point-group operations of the B-LM (i.e., of $G_b$) are part of $G_N$.

We consider $N\ge n_p$.
In this case, $G_N$ is a subgroup of $G_b$ because any operation of the ML-LM must also be an operation of the B-LM for a MDO polytype.
For $N<n_p$, this is not always true as we discuss in Appendix \ref{app:PG-few-layers}, and so this case requires an independent treatment.
We note that $n_p$ is either 1 or 2 for the most studied LMs, such as black phosphorous and rhombohedral graphite ($n_p=1$) or MoS$_2$, hBN, and Bernal-stacked graphite ($n_p=2$).
Since the modes plotted in the fan diagrams (i.e., relative rigid oscillations of the layers) only exist for $N\ge2$, the condition $N\ge n_p$ is thus not a strong limitation.

We first consider non-LOC operations.
Non-LOC $\sigma$ operations ($\sigma-\tau$) can never be symmetries of a finite ML-LM, because they map each layer with stacking index $\ell$ onto the layer with stacking index $\ell+1$.
For $\lambda$ non-LOC ($\lambda-\tau$) symmetries, these non-LOC layer-invariant operations form a group\cite{Scheuschner2015} that we call the layer-invariant point group, $G_I$, which is a subgroup of $G_b$.
Since all elements of $G_I$ leave each layer invariant individually (i.e., they map each layer onto itself\cite{Scheuschner2015}), they are also symmetry operations of the ML for any $N$.
Thus, $G_I$ is a subgroup of $G_N$.
Given a B-LM space group, obtaining $G_I$ amounts to considering all B-LM symmetry operations that are non-LOC, take only their rotational part, and check which point group they form.
$G_I$ for all space groups and settings is in Table \ref{tab:PGevolution}.

To obtain the complete $G_N$, we have to complement $G_I$ with all LOC ($\rho$) operations of the ML-LM, which are a subset of the LOC operations of the B-LM.
For Category II, no LOC operations exist in B-LM, see Table \ref{tab:operations-reference-per-category}.
Therefore, there are no additional operations to consider and $G_N=G_I$, independently of $N$ and $n_c$.
For Categories I and III, LOC operations exist, and we need to select the B-LM LOC operations compatible with a finite ML-LM.

We focus our attention on Category I because, as explained in Appendix \ref{app:grouping-catIII}, Category III can be considered as a special case of Category I for the determination of the point group.
For a ML-LM with $N$ layers, if an inversion centre for a LOC operation exists, this must be the layer plane of the central layer if $N$ is odd (e.g., layer with stacking index 3 if $N=5$), or the middle plane between the two central layer planes if $N$ is even.
Therefore, the ML-LM point group $G_N$ will be obtained complementing $G_I$ with only those LOC operations that have inversion planes respectively on a layer plane or on a middle plane.

For Category I there is always at least one operation with such a plane.
As explained in Appendix~\ref{app:grouping-catI}, if $n_c$ is odd, any LOC can be both considered as having a symmetry on a layer plane or (with a different fractional translation) on a middle plane, so all of them can be included when computing $G_N$.
Instead, for even $n_c$ (see Appendix~\ref{app:grouping-catI}), LOCs can either have inversion on a layer plane, or on a middle plane.
Depending on the parity of $N$, two point groups might alternate, corresponding to which set of LOCs is compatible with $N$.

Table \ref{tab:PGevolution} reports the complete set of possible ML-LM point groups $G_N$ for each Hall setting and for $n_c=1,\ldots,6$.
Table \ref{tab:PGevolution} often gives 2 symbols ($\diagup$ or $\times$) instead of one or two possible point groups $G_N$.
These indicate cases in which it is impossible to create a ML in that setting and with the specified $n_c$ in the conventional cell.
The meaning of the two symbols is explained in the Table caption and, in detail, in Appendix \ref{app:grouping-catI}.

We now illustrate with a few examples how to use Table \ref{tab:PGevolution} (noting that the online tool presented in Sec.~\ref{sec:tool} performs the symmetry analysis automatically without the need to check the table).

Given a LM, we first need to identify its layers, and determine in which category of Fig.~\ref{fig:MDO-categories} the materials falls in, depending on the symmetry of the 1L-LM.

Let us start with an example for Category I.
If we consider MoS$_2$ (in its 2H phase), hBN, or Bernal graphite, in all these cases the 1L-LM is non-polar (there is a symmetry operation that flips it upside down), so they belong to Category I, and the bulk space group is $P6_3/m2/m2/c$ (194), with a single choice of Hall number (488).$n_c=2$ in all these cases (see, e.g., Fig.~\ref{fig:MDO-categories}a).
Then, Table \ref{tab:PGevolution} shows that the possible ML-LM point groups are $\bar6$m$2$ and $\bar3$m.
$\bar6$m$2$ is for odd $N$ (which does not have a centre of symmetry) while $\bar3$m occurs for even $N$ ($\bar3$m has a centre of symmetry).
We emphasise here again the assumption $N\geq n_c$.
For graphene ($N=1$) the point group is $6/mmm$, i.e., neither of the two point groups occurring for $N\geq 2$ (for the reasons explained in Appendix~\ref{app:PG-few-layers}).
This can be understood considering that graphene has an additional centre of symmetry, that disappears in the graphite stacking for any odd $N>1$.
\begin{figure}
\centerline{\includegraphics[width=90mm]{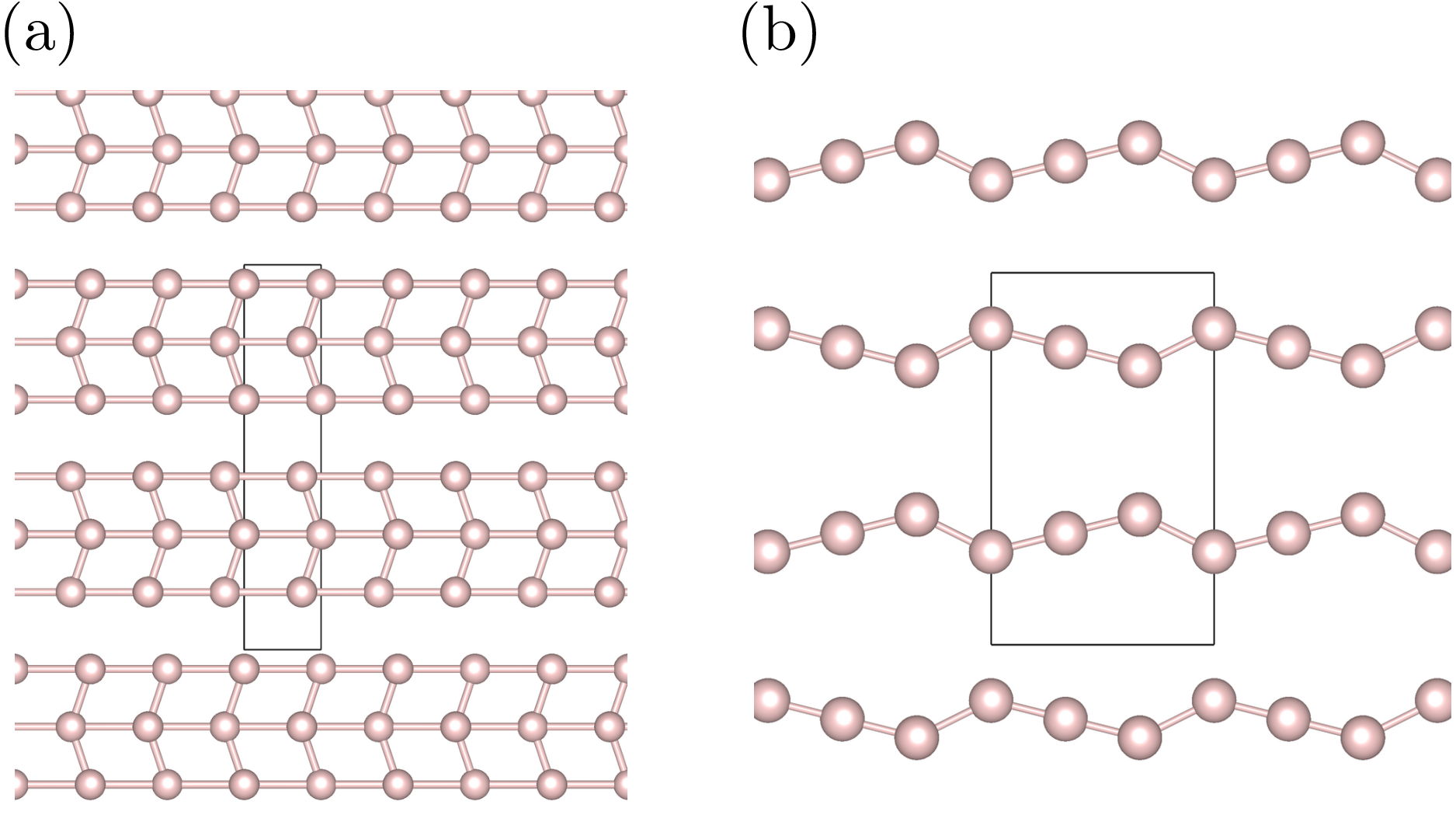}}
\caption{\label{fig:alternation-examples}
Two fictitious crystals with same B-LM space group (51, Hall number 242, Hall symbol P2/c2/m$2_1$/m) and $n_c=2$ (with an orthorhombic unit cell and translational invariance in the $y$ direction orthogonal to the page).
(a) 1L has mirror symmetry but no inversion.
The alternation of point groups for ML is 2/m for even $N$, mm2 for odd $N$.
(b) 1L has inversion symmetry but no mirror plane.
The alternation of point groups for ML is mm2 for even $N$, 2/m for odd $N$}
\end{figure}

From our analysis it is only possible to identify the set of possible $G_{N}$ given the Hall number and $n_c$.
To make a specific assignment for odd and even $N$, as in the above example, it is necessary to know the 1L symmetries.
To illustrate this, Figs.~\ref{fig:alternation-examples}a,b show two fictitious crystals with the same B-LM space group (51, Hall number 242, Hall symbol P2/c2/m2$_1$/m) and $n_c=2$.
From Table~\ref{tab:PGevolution} the 2 possibilities for $G_N$ are i) 2/m (having inversion) or ii) mm2 (not having inversion).
In both cases the B-LM has both inversion symmetry and horizontal mirror symmetry, with a corresponding B-LM point group $G_b = \mathrm{mmm}$.
However, 1Ls have either horizontal reflection symmetry only (Fig.~\ref{fig:alternation-examples}a) or inversion symmetry only (Fig.~\ref{fig:alternation-examples}b).
As a result, the inversion and mirror LOC operations have different centres in B-LMs, with the inversion centre on middle (layer) planes for Fig.~\ref{fig:alternation-examples}a (Fig.~\ref{fig:alternation-examples}b), and horizontal mirror symmetry on layer (middle) planes for Fig.~\ref{fig:alternation-examples}a (Fig.~\ref{fig:alternation-examples}b).
Since symmetries from middle planes are selected for even $N$ and those from layer planes for odd $N$, for Fig.~\ref{fig:alternation-examples}a the assignment is 2/m for even $N$ and mm2 for odd $N$.
The opposite holds for Fig.~\ref{fig:alternation-examples}b.

We now consider some examples from Categories II, III.
BiTeCl (Fig.~\ref{fig:MDO-categories}b) has B-LM space group P6$_3$mc (186, Hall number 480).
Since each layer is polar, and all layers have the same polarity (Category II), the point group of any ML-BiTeCl will be $G_I=3$m (see Table~\ref{tab:PGevolution}) as explained in Appendix~\ref{app:grouping-catI}.
For Bi$_2$TeI (Fig.~\ref{fig:MDO-categories}c), instead, the B-LM space group is C2/m (12, Hall number 63).
Since it belongs to Category III, the point group of any ML-Bi$_2$TeI with odd $N$ will be $G_I=\text{m}$.
If instead $N$ is even, then we need to check in Table~\ref{tab:PGevolution} the column for $n_c=1$ (as explained in Appendix~\ref{app:grouping-catIII}: Table~\ref{tab:PGevolution} can be used but interpreting $n_c$ in the table as the number of layer pairs, i.e., half of the number of layers in the bulk conventional cell), so that the resulting point group is 2/m, independent of the termination of the ML-Bi$_2$TeI.
Since, as we discussed earlier, some entries in Table~\ref{tab:PGevolution} have two possible values, this implies that for some space groups we might have an alternation of point groups $G_N$ for $N$ multiple of 4 or multiple of 2 only, and the specific point group taken will depend on the termination of the finite ML.
\subsection{Computing normal modes}
\label{sec:normal-modes}
In a fan diagram, we focus only on vibrational modes associated with a rigid relative motion of the layers, that are typically those with lower energy $< 100$~cm$^{-1}$.

The simplest approximation\cite{Tan2012} is to model the ML-LM as a finite linear chain model of masses with a force constant $K$ between them, which might depend on the direction of motion.
This model is often able to capture the qualitative behaviour of the frequency of the modes as a function of $N$, but it might not be able to predict accurately the frequencies or the coupling between C and LB modes in some systems.
Extensions of this model have been proposed to include further neighbours\cite{Wu2015} or intralayer coupling\cite{Wieting1973}, e.g., in the case of MoS$_2$, where a diatomic chain model has been derived\cite{Zha2013} to take into account the two types of atoms in the system (Mo and S).

Since layers are held together by van-der-Waals forces (typically much weaker than the chemical bonds between atoms in a layer), in the following we derive a more general tensorial model under the following two assumptions: 1) layers move as rigid units, i.e., the atomic displacements $\mathbf u(\ell)$ depend only on the stacking index $\ell$; 2) we include only first neighbour interactions between layers.
These two assumptions are typically very good in most ML-LMs\cite{Liang_review_2017,Zhang_review_2016,Ji_review_2016}.
Thus, even when they break, such as at the interface between twisted graphene multilayers\cite{Wu2015}, the predictions of our model are still useful to qualitatively interpret experimental data.

Under these assumptions, the equation of motion can be then written as:
\begin{align}\label{eq:full1Dmodel}
M \ddot{u}_{\alpha}(\ell) =  \sum_{\beta} \bigg\{ & K^{(\ell)}_{\alpha\beta} \left[u_{\beta}(\ell+1) - u_{\beta}(\ell)\right]   +\notag \\
& K^{(\ell-1)}_{\alpha\beta} \left[u_{\beta}(\ell-1) - u_{\beta}(\ell)\right]\bigg\},
\end{align}
where $M$ is the 1L total mass per unit cell, $\alpha$ and $\beta$ are Cartesian directions, and $K_{\alpha\beta}^{(\ell)}$ is the (tensorial) force constant between layer $\ell$ and $(\ell+1)$.
Eq.~\eqref{eq:full1Dmodel} is valid for B-LMs when periodic boundary conditions are applied, $u_{\beta}(\ell=n_c+1) =u_{\beta}(\ell=1)$, and for finite ML-LMs when all $u_{\beta}(\ell)$ terms for $\ell < 0$ or $\ell > N$ are removed.

The $K^{(\ell)}_{\alpha\beta}$ tensor, describing the interaction between two adjacent layers, can be different for each pair of layers.
For Category III of Fig.~\ref{fig:MDO-categories}, there are two types of interfaces that alternate: one set having Te atoms facing each other, the other having Bi atoms, and the corresponding force constants will thus be different.
Even for Categories I, II, where all layers and layer interfaces are identical, the matrices for different interfaces between layers $\ell$ and $\ell+1$ can differ, e.g., because an interface is obtained from the previous one by a rotation along the vertical axis, or some other symmetry operation (e.g., as for MoS$_2$ and BiTeCl, see Fig.~\ref{fig:MDO-categories}a,b).
In these cases, the matrices are related between them by the coincidence operation bringing one layer onto the next one, and we can write $K^{(\ell + 1)}=RK^{(\ell)}R^{-1}=R^{\ell -1}KR^{-\ell+1}$, with $R$ being the rotational part of the coincidence operation, and $K=K^{(1)}$ the interlayer force constant between the first and the second layer of the stack.
For Category III, $K^{(\ell)}$ can be generated in an analogous way starting from one of the two matrices $K^{(1)}$ and $K^{(2)}$, depending on the parity of $\ell$.
We thus expect in general not a single  $K_{\alpha\beta}^{(\ell)}$,  but a set of interlayer force-constant matrices, depending on a few parameters.

In the online tool described in Sec.~\ref{sec:tool} we apply and solve numerically Eq.~\eqref{eq:full1Dmodel}, and so we use the appropriately transformed $K^{(\ell)}$ for each layer.
However, in order to get a qualitative understanding of the frequencies, their degeneracies, and their interpretation as C or LB modes, we summarise here the analytical results when there is a single $K_{\alpha\beta}$ for all layer pairs (i.e., we are in Categories I or II, and the operation $R$ commutes with $K^{(1)}_{\alpha\beta}$, so that all $K$ matrices turn out to be identical).
This is the case, for instance, of MoS$_2$ or hBN.

Since $K_{\alpha\beta}$ is symmetrical, it can be diagonalised with eigenvalues $k_1$, $k_2$ and $k_3$.
Then, one can solve the equation of motion to obtain $3N$ solutions (for a ML with $N$ layers) obtaining\cite{Tan2012,Zhao2013}:
\begin{equation}\label{eq:nmodes}
u^{(\nu,n)}_{\beta}(\ell,t) = V_{\beta\nu} \cos\left[\frac{(n-1)(2\ell-1)\pi}{2N}\right] e^{i\omega^{(\nu,n)} t},
\end{equation}
where $V_{\beta\nu}$ are the eigenvectors of $K_{\alpha\beta}$ and $\nu=1,2,3$ denotes 3 branches (of $N$ modes each, indexed by $n = 1,\dots,N$).
The corresponding vibrational frequencies are given by:
\begin{align}
\omega^{(\nu,n)} &= \sqrt{\frac{2k_\nu}{M} \left[1-\cos\left(\frac{(n-1)\pi}{N}\right)\right]}= \notag \\
                 &= 2\sqrt{\frac{k_\nu}{M}} \sin\left(\frac{(n-1)\pi}{2N}\right).\label{eq:frequencies-chain}
\end{align}
The oscillation direction in each branch $\nu$ coincides with one of the principal directions of the symmetric tensor, identified by the eigenvector $V_{\beta\nu}$.
In order to define C and LB modes, corresponding respectively to oscillations parallel to the layers (in the $xy$ plane) and out-of-plane (along $z$), the $K$ matrix must be block-diagonal, with a $2\times2$ block for the C modes and a $1\times1$ element for the LB block.
In this case, we can then define if $\nu$ is a C or LB mode.
We note that the frequency of the highest C mode in a ML with $N$ layers is also called Pos(C)$_N=\frac{\omega^{(\text{C},N)}}{2\pi c}=\frac{1}{\pi c}\sqrt{\frac{k_\nu}{M}} \cos\left(\frac{\pi}{2N}\right)$, when expressed as a wavelength (with $c$ being the speed of light) and similarly Pos(LBM) refers to the highest LB mode.

In general, however, the $K$ matrix does not have such block form, and then the in-plane and out-of-plane vibrations are not decoupled, meaning that a distinction between LB and C modes is not possible, e.g., in the case of WTe$_2$ (see also Appendix~\ref{app:C-LB-separation-monoclinic} for an in-depth discussion on the separation of C and LB modes depending on the symmetry).

Since $K$ describes the interaction between adjacent layers, its tensorial form (which elements are zero, which elements are equal to each other) depends on the crystal system\cite{Nye} of the 2L-LM obtained by isolating the two layers, which is directly obtained from its point group.
\begin{figure}
\centerline{\includegraphics[width=80mm]{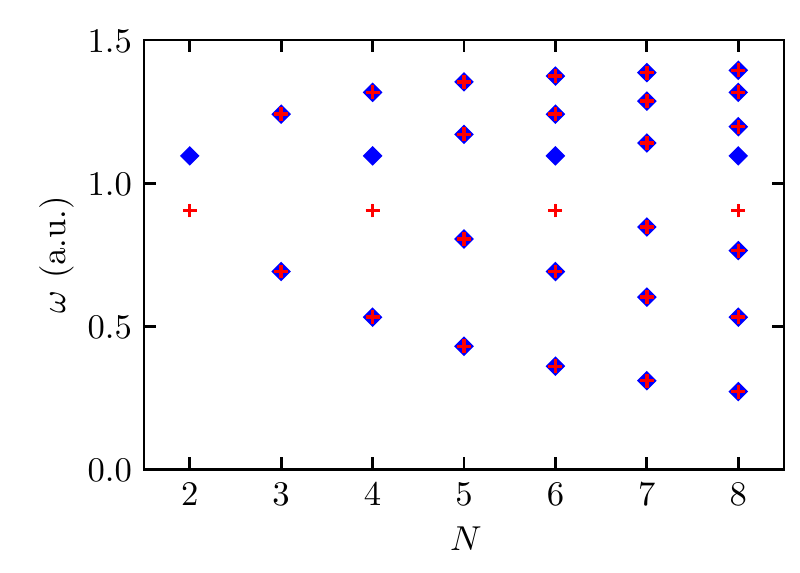}}
\caption{\label{fig:symbreak}C-modes fan diagram for ZnCl$_2$ obtained by solving Eq.~\eqref{eq:full1Dmodel}.
The phonon frequency is normalised to the mean of the two frequencies for $N=2$.
Red plus signs and blue diamonds denote C modes along the $x$ and $y$ in-plane directions, respectively.
Both B-ZnCl$_2$ and ML-ZnCl$_2$ with odd $N$ have tetragonal symmetry, and the two C modes are always degenerate.
However, ML-ZnCl$_2$ with even $N$ have a reduced orthorhombic symmetry, which removes the degeneracy between some of the C modes}
\end{figure}

All 7 possible cases\cite{Nye} (skipping cubic systems that, as discussed in Sec.~\ref{sec:multilayer-pointgroup}, are not compatible with a layered structure) are reported in Table~\ref{tab:K-tensor-shape}.
For 2L-LMs, we have that for trigonal, hexagonal, and tetragonal systems a distinction between LB and C modes is possible and the C branches are degenerate.
For orthorhombic systems, LB and C modes can be still defined, but the degeneracy of the two C branches is now lifted.
For monoclinic systems, we can distinguish 2 cases: (i) in-plane monoclinic unique axis, for which we can identify one pure C branch, while the other two branches are mixed (no pure LB mode can be defined, e.g., WTe$_2$); (ii) unique axis along the stacking direction, for which we can distinguish LB and C modes, even though the C polarisation has no specific orientation with respect to the crystal axes.
For triclinic systems, there is no symmetry constraint, therefore a distinction between LB and C modes is not possible (although there might still be a mode mostly polarised orthogonally to the layers, i.e., with a large LB character; this could happen for instance in LMHs).

\begin{table}[t]
\caption{\label{tab:K-tensor-shape}Components of the $K^{(\ell)}_{\alpha\beta}$ force-constants tensor (from general symmetry considerations, see Ref.~\onlinecite{Nye}) depending on the crystal system of the corresponding 2L-LM formed by layers $\ell$ and $(\ell+1)$.
The stacking direction is $z$.
Non-zero components are indicated, and components that are equal are indicated with the same name.
For monoclinic systems, we need to distinguish the case where the unique axis is in plane (here arbitrarily chosen as $y$) or along $z$}
    \begin{tabular}{cc}
        \begin{minipage}{4cm}
            tetragonal,\\
            hexagonal or\\
            trigonal
        \end{minipage} &
        $\begin{pmatrix}
            xx & 0 & 0 \\
            0 & xx & 0 \\
            0 & 0 & zz
        \end{pmatrix}$ \\
        orthorhombic &
        $\begin{pmatrix}
            xx & 0 & 0 \\
            0 & yy & 0 \\
            0 & 0 & zz
        \end{pmatrix}$ \\
        \begin{minipage}{4cm}
            monoclinic ($y$):\\
            in-plane\\
            unique axis
        \end{minipage} &
        $\begin{pmatrix}
            xx & 0 & xz \\
            0 & yy & 0 \\
            xz & 0 & zz
        \end{pmatrix}$ \\
        \begin{minipage}{4cm}
            monoclinic ($z$):\\
            out-of-plane\\
            unique axis
        \end{minipage} &
        $\begin{pmatrix}
            xx & xy & 0 \\
            xy & yy & 0 \\
            0 & 0 & zz
        \end{pmatrix}$ \\
        triclinic &
        $\begin{pmatrix}
            xx & xy & xz \\
            xy & yy & yz \\
            xz & yz & zx
        \end{pmatrix}$ \\
    \end{tabular}
\end{table}

For a ML-LM with $N>2$, while the optical activity (discussed in the next section) and the degeneracies do depend only on the point group of the ML-LM (as obtained in Sec.~\ref{sec:multilayer-pointgroup}), in general the considerations of the previous paragraph on when we can define pure C and LB modes cannot be naively applied (see Appendix~\ref{app:C-LB-separation-monoclinic} for more details and an example).

Since not only $G_N$  often differs from $G_b$ (e.g., in MoS$_2$, hBN), but it might also belong to another crystal system, the degeneracies of the modes might be different in B-LM and ML-LM.
For instance (see also Appendix~\ref{app:PG-few-layers}), B-WTe$_2$ is orthorhombic (space group Pmn2$_1$, Hall number 155), but for all $N$ the ML-WTe$_2$  point group is always m, a monoclinic point group.
In other cases, this occurs only for some values of $N$, like for ZnCl$_2$ (tetragonal bulk, Hall number 420), where the ML-LM point group is $\bar 4$2m (tetragonal) for odd $N$, but is mmm (orthorhombic) for even $N$.
The degeneracies vary with $N$ in this case, as illustrated in Fig.~\ref{fig:symbreak}, and can be used as an additional fingerprint of the material.

\subsection{Optical activity of a multilayer}
\label{sec:optical-activity}
Once the point group of a ML-LM and its normal modes (frequencies and eigenvectors) are known (thanks to the results of the previous sections), one can assess its Raman or IR activity.
This can be done by projecting the normal modes onto the different irreducible representations of the point group (listed in standard crystallography references\cite{Cracknell,Altmann1994,Aroyo2011}) to understand which one they belong to.
In particular, apart from accidental degeneracies, a normal mode belongs to one and only one irreducible representation\cite{Dresselhaus}, provided that pairs of complex representations that are conjugate of each other are grouped together because of time-reversal symmetry.
Thus, the following expectation value will be one for the irreducible representation $\gamma$, with characters $\chi^{(\gamma)}(g)$, to which the normal mode $(\nu, n)$ belongs, and will be zero for all other representations\cite{Dresselhaus}:
\begin{equation}
p_{\gamma}(\nu, n) = \frac{d_{\gamma}}{h} \sum_{g\in G} \left[\chi^{(\gamma)}(g)\right]^{*} \left.{\mathbf U}^{(\nu,n)}\right.^{\dag} \hat O_{g} {\mathbf U}^{(\nu,n)},\label{eq:optical-activity}
\end{equation}
where ${\mathbf U}^{(\nu,n)}$ is a vector collecting the displacements ${\mathbf u}^{(\nu,n)}(\ell)$ of the layers obtained by solving Eq.~\eqref{eq:nmodes}, $d_{\gamma}$ is the dimension of the representation, $h$ the order of the point group, and $\hat O_{g}$ the operator associated with the symmetry element $g$ (all these are tabulated for all point groups).
From the knowledge of the representation $\gamma$ for which $p_{\gamma}(\nu, n)=1$, we can determine if the mode is Raman and/or IR active depending on whether the representation transforms as the components of a vector ($x$, $y$, $z$) or of a quadratic form ($x^2$, $y^2$, $xz$, \ldots), respectively.
Additionally, if there exists at least one quadratic form associated with $\gamma$ that does not involve the $z$ coordinate, the mode should also be visible in a back-scattering Raman geometry, as the light polarisation vector in a back-scattering experiment with light propagating along $z$ cannot have a $z$ component.

To showcase the application of the method, Tables~\ref{tab:layerbreathing},\ref{tab:shear} report the results obtained for all point groups, when a single force-constant tensor is sufficient, so that the analytical expressions of Sec.~\ref{sec:normal-modes} can be adopted.
In particular, for each mode of a $N$-layer ML-LM with point group $G_N$, we indicate the irreducible representation to which it belongs together with its IR/Raman activity, and whether the mode visible in a Raman spectroscopy experiment with a back-scattering geometry.
The full analysis for any input LM, also when more than one force-constant tensor is needed, is performed by our online tool (see Sec.~\ref{sec:tool}).

\section{Results}
\label{sec:results}
We now show with a few examples how to use this approach to reconstruct the fan diagram and the pattern of modes detectable in IR or Raman spectroscopy.

Let us start from the case of MoS$_2$ and black phosphorous (BP).
As discussed in Sec.~\ref{sec:multilayer-pointgroup}, the ML point group is $\bar6$m2 for odd $N$ and $\bar3$m for even $N$.

\begin{figure*}
\begin{tabular}{cc}
\includegraphics{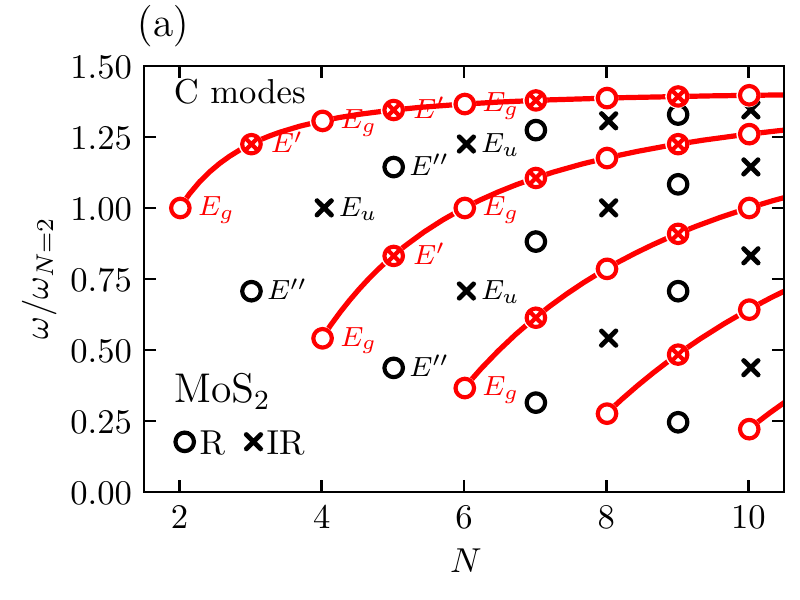} & \includegraphics{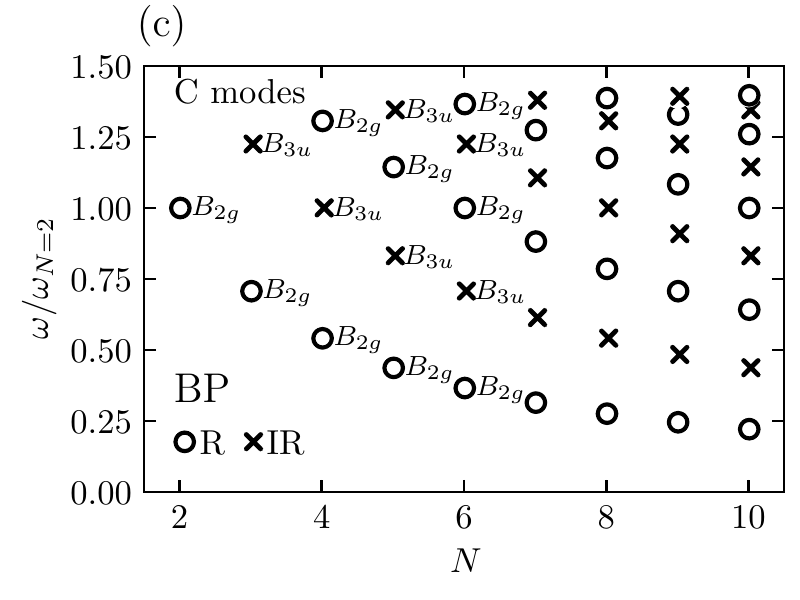}  \\
\includegraphics{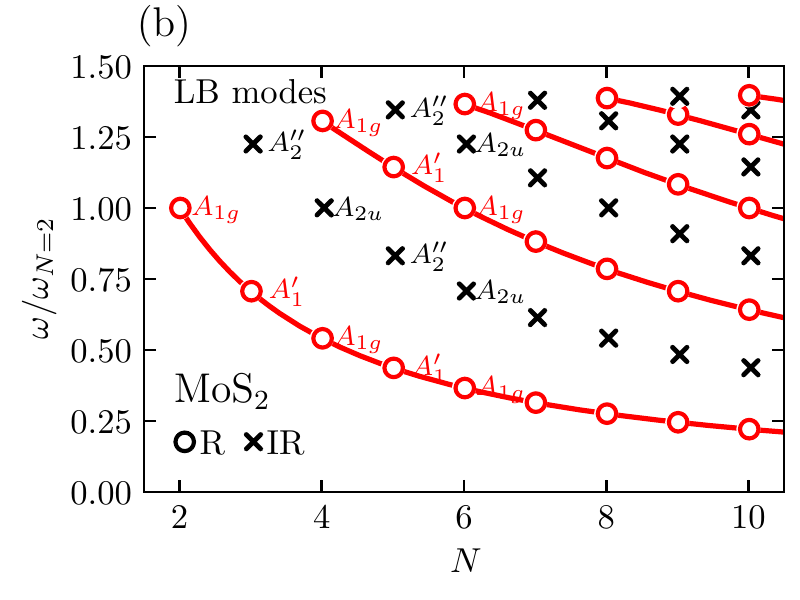} & \includegraphics{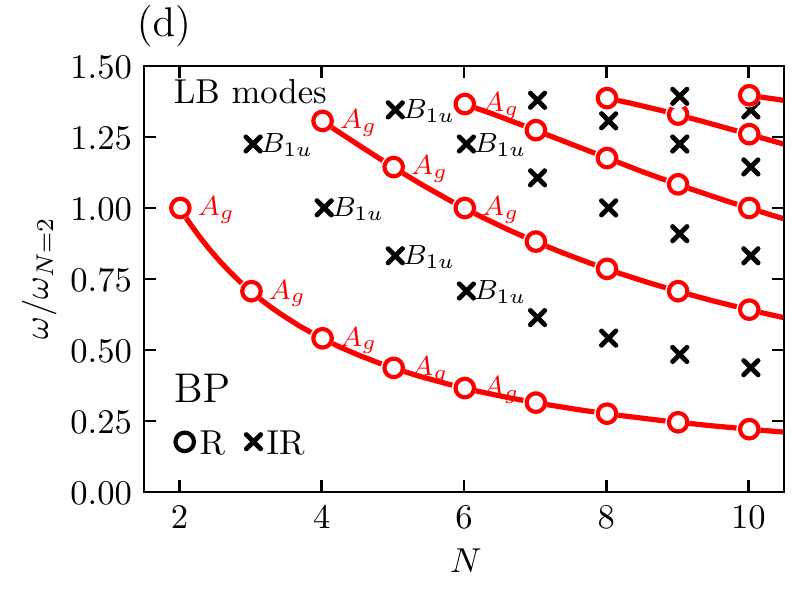} \\
\end{tabular}
\caption{Fan diagram for C modes (panels a, c) and LB modes (panels b, d) for ML-MoS$_2$ (panels a, b) and ML-BP (panels c, d).
Open circles indicate Raman-active modes, while crosses indicate IR-active modes.
Red symbols denote Raman-active modes that are detectable in a back-scattering geometry.
For each mode with $N\leq6$, the corresponding irreducible representation of the point group is reported (note that
in the case of ML-BP two non-degenerate sets of C modes exist but only one of them is reported with the corresponding irreducible-representation names).
Red lines are guides to the eye, following the pattern of Raman-active modes visible in back-scattering.
The frequency $\omega$ on the $y$ axis is normalised to the frequency $\omega_{N=2}$ of the corresponding mode in 2L.
This frequency is different between C and LB modes
\label{fig:MoS2-BP}}
\end{figure*}

Figs.~\ref{fig:MoS2-BP}a,b plot the fan diagrams for the C and LB modes of ML-MoS$_2$  as a function of $N$, where the assignment of the modes is obtained by considering the appropriate entries in Tables~\ref{tab:layerbreathing},\ref{tab:shear}, and reproduce the experiments in Refs.~\onlinecite{Zha2013,Zhao2013}.

We then consider ML-BP whose bulk space group is A2$_1$/b2/m2/a (space group 64, Hall number 306 when considering the shortest in-plane vector along the second axis), $n_c=2$ and the corresponding ML-BP point group is mmm both for even and odd $N$.
Figs.~\ref{fig:MoS2-BP}c,\ref{fig:MoS2-BP}d report the corresponding fan diagrams, reproducing the experiments of Ref.~\onlinecite{Luo2015bp}.
We note that, in this case, C modes are not visible in back-scattering, consistent with Ref.~\onlinecite{Luo2015bp}.

\begin{figure*}
\begin{tabular}{cc}
\includegraphics{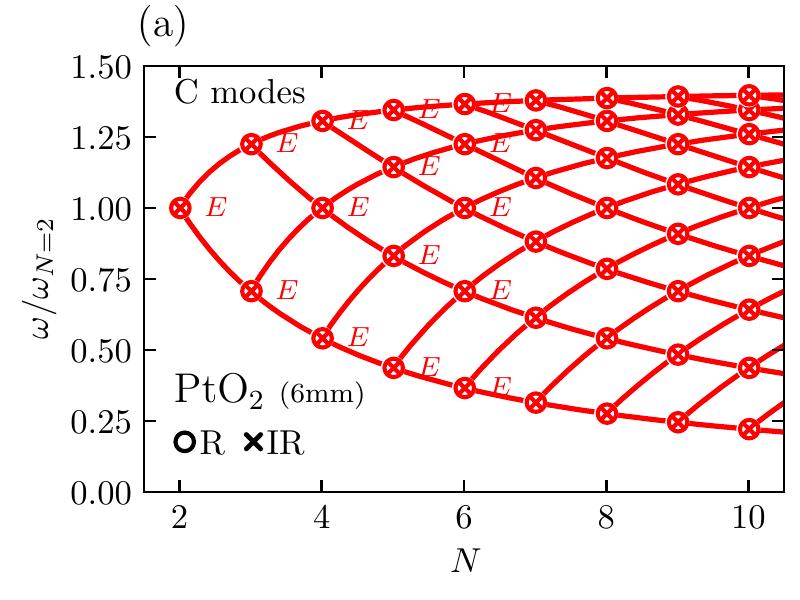} &
\includegraphics{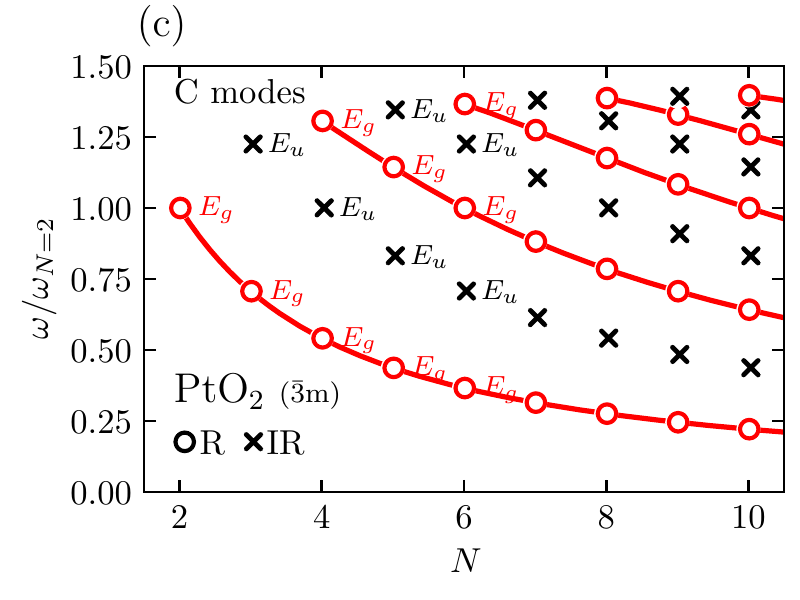} \\
\includegraphics{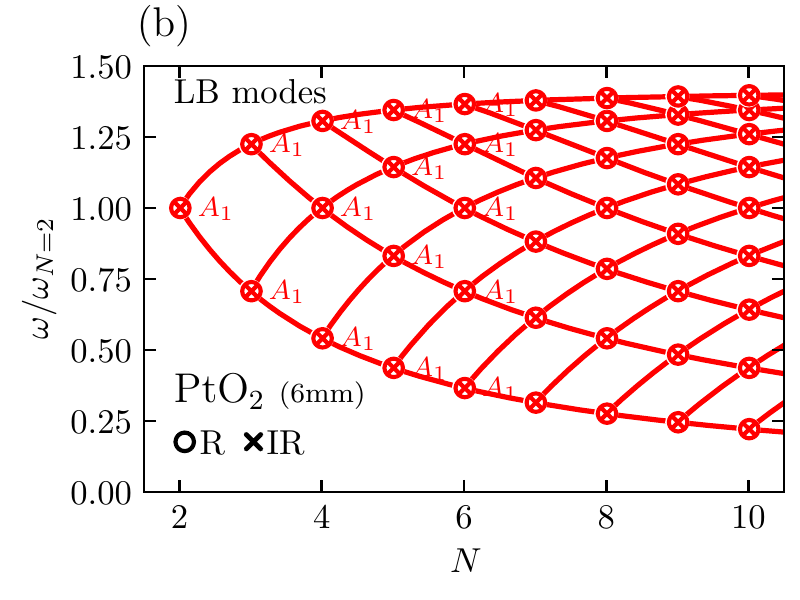} &
\includegraphics{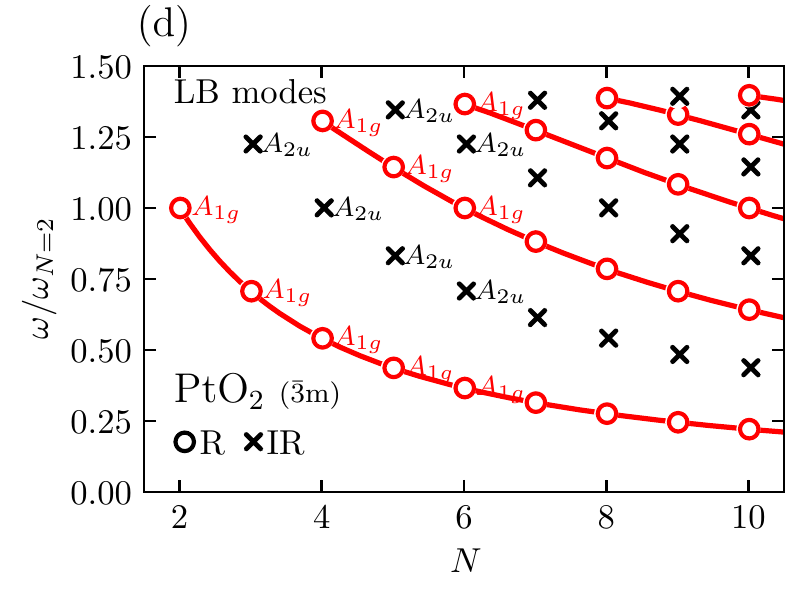} \\
\end{tabular}
\caption{Fan diagram for C modes (panels a, c) and LM modes (panels b, d) of ML PtO$_2$ for two different bulk allotropes with space group P$6_3$mc (panels a, b, point group 6mm) and P$\bar3$m1 (panels c, d, point group $\bar3$m).
See Fig.~\ref{fig:MoS2-BP} for the meaning of the symbols and the colours
\label{fig:PtO2}}
\end{figure*}

As a further example, Fig.~\ref{fig:PtO2} shows the fan diagram of PtO$_2$, which can crystallise in at least two different allotropes differing only in the layer-stacking sequence\cite{Hoekstra1971}.
One phase has space group P$6_3$mc (space group 186, Hall number 480) with $n_c=2$; the other has $n_c=1$ and space group P$\bar3$m1 (space group 164, Hall number 456).
In the first case, the ML-PtO$_2$ point group is always 3m as reported in Table~\ref{tab:PGevolution}, so that all C and LB modes are Raman-active in back-scattering (see Tables \ref{tab:layerbreathing},\ref{tab:shear}).
In the second case, the point group is $\bar3$m for every $N$, and the pattern of Raman-active modes is in Figs.~\ref{fig:PtO2}c,d.
Since the pattern is different from the first phase, this implies that the pattern of Raman-active modes detectable in back-scattering can be used as a fingerprint to recognise the stacking sequence and symmetry properties of a given ML-PtO$_2$.
\section{Online tool}
\label{sec:tool}
\begin{figure}[tb]
\centerline{\includegraphics[width=\linewidth]{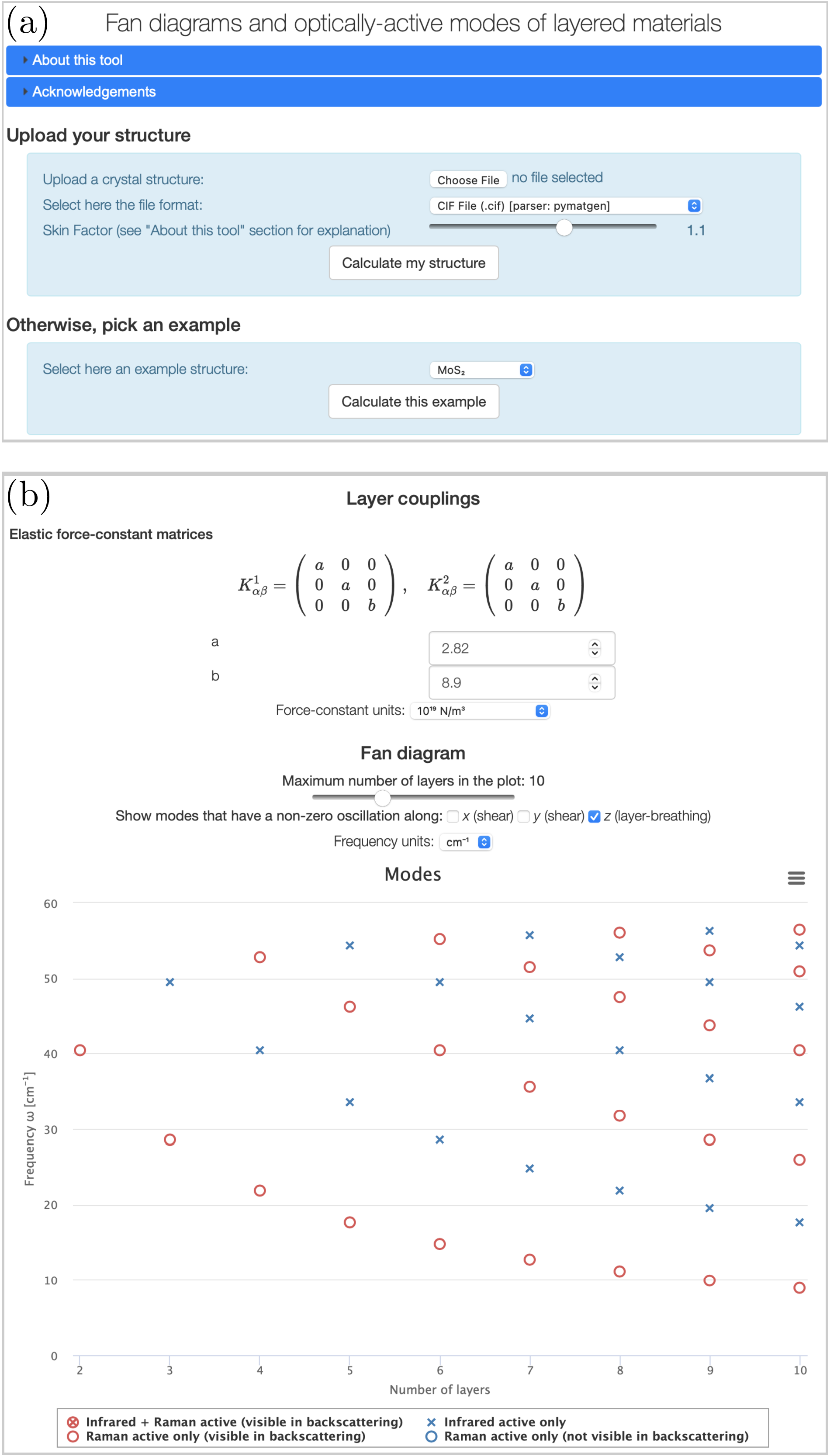}}
\caption{\label{fig:tool}Screenshots of the online tool implementing the algorithms of this paper, available on the Materials Cloud~\cite{Talirz2020} Work/Tools section.
(a) Selection page, where it is possible to upload a structure in a number of common formats, or to select an example.
(b) Part of the output page showing the resulting fan diagram for a material, in this case for MoS$_2$, where the option to display only LB modes has been selected.
The output page of the tool actually displays much more information, like visualisations of the crystal structure of B-LM and of the layers, the coincidence operation of the ML-LM, and the symmetry analysis for the B-LM, 1-LM, and ML-LM for all possible values of $N$
}
\end{figure}

In order to make the algorithm described in this paper readily available to any researcher, we implemented it in an online web tool, published on the Materials Cloud web platform\cite{Talirz2020} at the address \url{https://materialscloud.org/work/tools/layer-raman-ir}.
This tool does not require any installation and works directly in the browser.
In the first selection page, shown in Fig.~\ref{fig:tool}a, the user can upload the bulk crystal structure of a LM in a number of common formats, leveraging the parsers implemented in the ASE\cite{ASE} and pymatgen\cite{pymatgen} libraries.
A ``skin factor'' parameter $f$ can also be selected to tune the bond-detection algorithm.
In particular, the tool considers two atoms A and B bonded if their distance is $<f\cdot(r_A + r_B)$, where $r_A$ and $r_B$ are the corresponding covalent atomic radii from Ref.~\onlinecite{Cordero2008}.
Alternatively, it is possible to choose among a few selected examples that we provide as demonstrators.

Once the bulk structure is selected or uploaded, the tool performs computations in the background and produces an output page.
It first computes the bonds and then detects the disconnected lower-dimensional components.
Once these are determined, it checks that all these components are two-dimensional and identical between them (using the pymatgen code\cite{pymatgen} and, in particular, the \texttt{structure\_matcher} module, to compare layers, check if they are identical within a numerical threshold, and determine which coincidence operation brings one onto the other).
It then rotates the whole structure so that the stacking axis is along $z$ and computes the coincidence operation between each pair of layers in the conventional cell, verifying that the system satisfies the hypotheses of this paper (same coincidence operation between any pair of consecutive layers) and assigning it to one of the three categories of Fig.~\ref{fig:MDO-categories}.
If any of the previous steps does not succeed, the tool displays a message informing that the structure does not satisfy the assumptions of this work.
After this geometry analysis, the tool determines the symmetry of B-LM and 2L-LM, thus the number and shape of the force-constant matrices.
Extending the assumptions used here to produce Tables~\ref{tab:layerbreathing},\ref{tab:shear}, the tool works also in the case in which the force constant $K$ and the rotational part of the coincidence operation $R$ do not commute, such as for instance in WTe$_2$ and ZnCl$_2$, where force-constant matrices between successive layer pairs are related by symmetry but are not identical.
The output page then includes relevant information on the structure (interactive visualisations of B-LM and of 1L-LM, information on the coincidence operation), and shows the independent components of the force-constant matrices.
An initial random value for these components is provided, chosen to be in the range of those typically occurring in LMs, but these can be changed interactively (e.g., to fit experimental data, or to use values obtained from first principles).
The tool then computes the corresponding fan diagram, including the optical activity for IR and Raman spectroscopy.
Multiple units are supported both for the force constants and for the phonon frequencies.
A screenshot of the resulting fan diagram as provided by the tool (including the section to select the force-constant parameters) is in Fig.~\ref{fig:tool}b.

\section{Conclusions}
We presented an approach to seamlessly predict the spectroscopic fingerprints of layered materials composed of repetitions of the same layer.
We explained how to obtain, using symmetry arguments, the point group of a finite multilayer, knowing the space group and the Hall setting of the bulk, and provided a table for all possible space groups and settings.
We derived the vibrational modes for any number of layers using a tensorial linear chain model.
We then exploited these results to associate each normal mode to a given irreducible representation of the point group of the multilayer, in order to assess the corresponding optical activity, and thus obtain the fan diagram and the pattern of modes that are detectable in IR and Raman spectroscopy.
We demonstrated with various examples that this can distinguish different stacking sequences of a given layered material and provides stringent conditions on the symmetry properties of multilayers.

We also provided an easy-to-use online web tool that allows users to upload a bulk multilayer system of their choice (accepting a variety of common crystal-structure formats) and performs all needed operations described in this paper to obtain and display interactively the corresponding fan diagram, even beyond some of the approximations used in this paper (like those used in Tables~\ref{tab:layerbreathing},\ref{tab:shear}).
The tool is available on the Materials Cloud web platform\cite{Talirz2020} at the address \url{https://materialscloud.org/work/tools/layer-raman-ir} and it is fully open-source (the code can be found at \url{https://github.com/epfl-theos/tool-layer-raman-ir}).
This will guide computational and experimental researchers interested in studying or interpreting fan diagrams of layered materials.

\section*{Acknowledgements}
We greatly acknowledge Radovan \v{C}ern\'y for useful discussions and Leopold Talirz for help in deploying the tool on the Materials Cloud.
We acknowledge funding from the MARVEL National Centre of Competence in Research of the Swiss National Science Foundation (SNSF) (grant agreement ID 51NF40-182892), from the European Centre of Excellence MaX ``Materials design at the Exascale'' (grant no. 824143), from the swissuniversities P-5 ``Materials Cloud'' project (grant agreement ID 182-008), from the EPFL Open Science Fund via the OSSCAR project, from the Graphene Flagship, ERC Grant Hetero2D, and EPSRC Grants EP/509K01711X/1, EP/K017144/1, EP/N010345/1, EP/M507799/5101 and EP/L016087/1, from the Italian Ministry for University and Research through the Levi-Montalcini program, and from SNSF through the Ambizione program (grant 174056).

\appendix
\section{Compatibility relations of fractional translations}
\label{app:compatibility}
A fractional translation $\bm \tau$ is a non-zero translation part of a space group transformation, that is by convention applied after the rotation matrix $R$, so that the coordinate transformation then reads:
\begin{equation}
\label{eq:transformation-fractional}
\mathbf r\to R\mathbf r + \bm \tau.
\end{equation}
Since we focus on LMs stacked along the $z$ axis, we consider only the $\tau_z$ component of the fractional translation vector.
In order for an operation with a non-zero fractional translation to be compatible with a LM with $n_c$ layers in the conventional cell, the product $n_c\cdot \tau_z$ must be an integer: e.g., if we consider a space group with a $3_1$ screw axis along $z$, it might be possible to construct a LM with this space group having $3, 6, \ldots{}$ layers in the B-LM conventional cell, but it is not possible to define a ML system having $n_c=1, 2, 4, 5, \ldots$
In Table~\ref{tab:PGevolution} we indicate therefore with a slash ($\diagup$) any space group that contains at least one incompatible operation for a given value of $n_c$.

Non-vanishing fractional translations (in the case of MDO polytypes) are therefore admissible only when $n_{c}=2$, 3, 4, 6.
This follows from the usual crystallographic conditions for which, for e.g., if we rotate a layer by an arbitrary angle, the next layer cannot be periodic with the same unit cell except for a few angles (see Chapters 1 and 2 of Ref.~\onlinecite{OKeeffe}).

\section{Grouping fractional translations of LOC operations: Category I}
\label{app:grouping-catI}
As discussed in the main text, in order to obtain $G_N$ of a ML-LM we need to identify the B-LM LOC operations compatible with a ML with a given $N$.
These, together with the elements of the layer-invariant point group $G_{I}$, will form the $G_N$ group that we seek for.

We now consider independently the 3 categories of Fig.~\ref{fig:MDO-categories}.
In Category II, there are no LOC operations, therefore $G_N=G_I$.
We focus in the rest of this Appendix on Category I and we show in the next Appendix that for Category III we can adapt the results of Category I.

We consider the subset of LOC operation of a given space group (and setting), defined as those that swap the orientation of the $z$ axis, i.e., where the third column of the rotation matrix $R$ is the vector $(0, 0, -1)$.
Focusing only on the third coordinate $z$ of a coordinate vector $\mathbf r$ and using Eq.~\eqref{eq:transformation-fractional}, the transformation will therefore read:
\begin{equation}
\label{eq:LOC-coordinate-transfomation-fractional}
z\to -z + \tau_z.
\end{equation}
Let us first fix the origin of our coordinate system by  setting the origin on  the inversion plane of the $i$-th LOC operation, which will then have no fractional translation along the vertical direction ($\tau^i_z=0$).

If we now choose another LOC operation, say the $j$-th, we might need to associate with it a non-zero fractional translation $\tau^j_z$ along the vertical direction.
In order to connect the coordinate of the inversion planes $\tilde z_j$ for this $j$-th LOC to its fractional translation $\tau^j_z$, we note that the $j$-th transformation can be equivalently interpreted as the combination of the following operations: 1) translating one inversion plane at $z=\tilde z_j$ to $z=0$ with a transformation $z\to z - \tilde z_j$; 2) applying the inversion transformation about the plane that is now at $z=0$, therefore changing the sign of the $z$ coordinate, so that the combined transformation reads $z\to -(z - \tilde z_j)$; 3) shifting back the inversion plane to its original position by adding $\tilde z_j$ to the third coordinate.
The total transformation is thus $z\to -(z - \tilde z_j) + \tilde z_j = -z + 2\tilde z_j$.
Comparing this with Eq.~\eqref{eq:LOC-coordinate-transfomation-fractional}, we obtain that the fractional translation $\tau^j_z$ must be $\tau^j_z=2\tilde z_j$.

As we discussed earlier (see Fig.~\ref{fig:MDO-categories}), for Category I inversion centres can only be on a layer plane or on a middle plane.
Having also chosen earlier the origin on one of these planes, the $\tilde z_j$ coordinate of any centre (in fractional coordinates) can thus only be at position at position $\tilde z_j=k/2n_c$, with $k\in\mathbb N$.
E.g., in the case of two layers A and B in the conventional cell ($n_c=2$), centres will be at position $\tilde z_j=0$ (on layer A), $\tilde z_j=1/4$ (between layer A and layer B), $\tilde z_j=1/2$ (on layer B) or $\tilde z_j=3/4$ (between layer B and layer A in the next unit cell).
Thus, fractional translations for any inversion plane can only assume values $\tau_z=k/n_c$, with $k \in \mathbb N$.

We can then use the information on the fractional translations to group all LOC operations in sets that share the same inversion plane(s), distinguishing in general those operations having $\tau_z=2h/n_c$ (with $h\in \mathbb N$) and thus inversion on a layer plane, from those having  $\tau_z=(2h+1)/n_c$, with inversion on a middle plane.
If $n_c$ is odd, the two sets are equivalent, i.e., each LOC transformation with inversion on a layer plane can be also written as an operation with inversion on a middle plane and a different fractional translation.
E.g., in the case $n_c=3$, one of the two groups would be $\{\tau_z=0, 2/3, 4/3, 6/3=2, \ldots\}$ and the second $\{\tau_z=1/3, 3/3=1, 5/3, 7/3, \ldots\}$.
Remembering that adding an integer to $\tau_z$ does not change the operation, we have that $4/3$ is equivalent to $1/3$, $2$ to $0$, $5/3$ to $2/3$ and so on, so that both sets coincide with $\{0, 1/3, 2/3\}$.
If $n_c$ is even, instead, there are two separate sets of fractional translations, giving rise to transformations having inversion either on layer planes or middle planes.
E.g., for $n=4$, one such set contains $\{\tau_z=0, 1/2\}$ and the other $\{\tau_z=1/4, 3/4\}$.

For each $\tau_z$ in one of these sets, we can construct a potential point group $G^{\tau_z}_N$ by adding to $G_I$ all LOC operations with fractional translation $\tau_z$.
In order to be consistent with our initial assumption of a layered structure with $n_{c}$ identical layers per cell and with the same relation between nearest layers (MDO polytypes), all the various possible point groups $G^{\tau_z}_N$ should be identical for all fractional translations belonging to the same set.
This stems from the fact that for Category I MDO polytypes all layer planes are equivalent, as well as all middle planes.
If this is not the case, we indicate it with a cross ($\times$) in Table~\ref{tab:PGevolution}.
E.g., in the case of space group $P\bar 1$  (Hall number 2) the bulk point group is $\bar 1$ while the layer-invariant point group $G_I$ is 1.
In the case $n_c=3$, considering LOC operations with fractional translation $\tau_z=0$ would add the $\bar1$ operation and would give rise to point group $G^{0}_N=\bar 1$.
However, considering operations with $\tau_z=1/3$ (or $\tau_z=2/3$) would give rise to a different point group $G^{1/3}_N=1$ ($G^{2/3}_N=1$), since $P\bar 1$ has no LOC operations with these fractional translations, and thus $G^{\tau_z}_N=G_I$ in this case.
These point groups ($1$ and $\bar 1$) are not the same and therefore we mark this case with a $\times$, indicating that it is not possible to construct a ML with symmetry $P\bar 1$ and $n_c=3$ identical layers with the same relation between each pair of them.

We summarise the results as follows: if $n_c$ is odd, we can either obtain a $\diagup$ or a $\times$, or there will be only one possible value for $G_N$, independent of the value of $N$.
When $n_c$ is even, the only difference is that in general there can be two possible choices for the point group $G_N$.
Which value is taken in the finite ML depends on the parity of $N$: the only LOCs compatible with a finite ML are those with symmetry plane at its centre (a middle plane for even $N$ or a layer plane for odd $N$).
Therefore in these cases the two possible point groups alternate as a function of $N$.

E.g., in the example of Fig.~\ref{fig:alternation-examples}, the Hall number is 242 (Hall symbol P2/c2/m2$_1$/m), $n_c=2$, and $G_I=\text{m}$.
Since we have a $2_1$ vertical axis, $n_c$ can only be even as discussed in Appendix~\ref{app:compatibility} and in Table~\ref{tab:PGevolution} there is a $\diagup$ for all odd $n_c$.
If we add LOC operations with a given fractional translation to $G_I$, we obtain point group $G_I=\text{m}$ for fractional translations 1/6, 1/4, 1/3, 2/3, 3/4 and 5/6 (since there is no additional LOC operation with these fractional translations).
We obtain instead 2/m for fractional translation 0, and mm2 for fractional translation 1/2.
Therefore, for $n_c=2$ we have two independent sets of fractional translations (\{0\} and \{1/2\}), and we thus obtain the two valid options for $G_N$: 2/m and mm2.
However, for $n_c=4$ (and similarly for larger even values of $n_c$) we obtain a $\times$, because one set of fractional translations \{0, 1/2\} (that must be equivalent for $n_c=4$) would instead contain two different point groups 2/m and mm2.

From pure symmetry considerations it is not possible to establish which of the two point groups takes place for odd or even $N$, as discussed in Fig.~\ref{fig:alternation-examples}, unless something is known for 1L.

\section{Grouping fractional translations of LOC operations: Category III}
\label{app:grouping-catIII}
If we limit to symmetry considerations (e.g., for the determination of the results of Table~\ref{tab:PGevolution}), we note that Category III is equivalent to Category I.
Indeed, if we consider a pair of adjacent layers in Category III, these together can be considered as a (now non-polar) ``layer'' of Category I.
In particular, the $\sigma-\rho$ plane between the pairs takes the role of the $\sigma-\rho$ middle plane of Category I, and the $\sigma-\rho$ plane between the layers of the pair takes the role of the $\lambda-\rho$ layer plane of Category I.
Note that there are two ways of pairing adjacent layers, and changing such choice swaps the role of middle planes and layer planes.

Intuitively, we can understand why these two categories are equivalent with the following \emph{gedankenexperiment}: if the chemical bonding between the two layers in a pair becomes stronger, without changing the atomic positions (and so without any change to the symmetry of the system), we will eventually end up considering both layers in the pair to be chemically bonded, therefore part of the same rigid layer.
In this case, then, we would have considered the system as belonging to category I.
Therefore, for the purpose of knowing the possible point groups, Table~\ref{tab:PGevolution} can still be used, with the caveat that now $n_c$ indicates the number of pairs of layers for category III.

We emphasise, however, that a separate treatment is needed when we consider the force constants between layers.
In this case, the strength of the chemical bonds matters in determining which layers can be considered as moving rigidly, and we need in general to consider two different sets of force constants for the various $\sigma-\rho$ planes of Category III.

In conclusion, for determining the point group of a ML with $N$ layers, there are now the following options:
\begin{itemize}
\item $N$ is odd.
In this case, on one of the two terminations there is only one layer in a pair.
The ML loses all LOC symmetries, and $G_N=G_I$.
\item $N$ is even.
We can then map this case to Category I, considering a system with $\tilde N=N/2$ pairs of layers as discussed above.
Depending on the parity of $n_c$ (which now indicates the number of pairs of layers in the conventional cell) we might have only one or two possibilities for the resulting point group $G_N$.
The termination of the finite ML will uniquely determine how to pair together adjacent layers.
\end{itemize}
Therefore, for Category III, there might be up to three different point group values as a function of $N$.

Finally, also ``dimerised'' systems with non-polar layers, where the interlayer distance alternates (A/B/A/B/\ldots), are still MDO polytypes and behave like those of Category III, and the symmetry plane of the $\sigma-\rho$ coincidence operation does not coincide with the layer plane.
We do not consider them explicitly here (and they are quite unlikely to occur in real ML systems) but the online tool is able to correctly consider also these and mark them as Category III.

\section{Point group of ML-LMs with $N<n_p$}
\label{app:PG-few-layers}
In the main text we have focused the discussion on the case $N \geq n_p$.
In this case, we can deduce $G_N$ starting from $G_b$ and removing the operations that are not valid in a ML-LM composed of $N$ layers.
The operations that remain form $G_N$, therefore $G_N$ is always a subgroup of $G_b$.

\begin{figure}[tbp]
\centerline{\includegraphics[width=90mm]{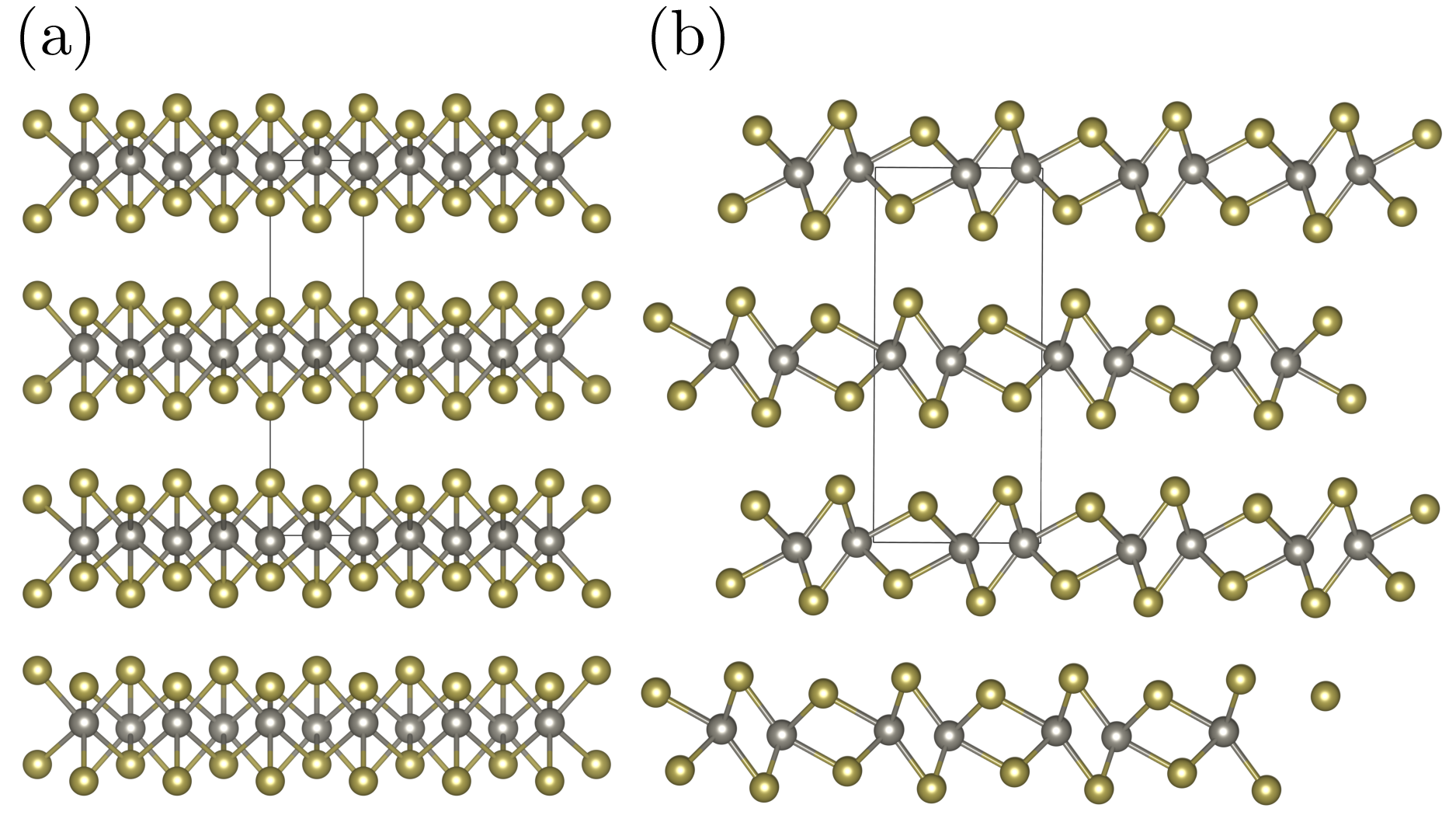}}
\caption{\label{fig:WTe2}ML-WTe$_2$ (COD\cite{COD} entry ID 2310355; grey: W, yellow: Te).
(a) Side view ($x-z$ projection).
(b) Side view ($y-z$ projection).
B-WTe$_2$ has space group Pmn2$_1$ (number 31) with two layers both in the conventional and the primitive unit cells ($n_c=n_p=2$; the unit cell is shown); $G_b$ is $mm2$ ($C_{2v}$).
With the given choice of axes the Hall setting is 155 (Pmn2$_1$), with a mirror plane orthogonal to $x$, a glide plane orthogonal to $y$, and a $2_1$ screw axis along $z$.
1L-WTe$_2$ has space group P2$_1$/m (with inversion symmetry and a $2_1$ screw axis along $x$), thus $G_{N=1}$ is 2/m ($C_{2h}$).
There is no group-subgroup relation between mm2 and 2/m.
From Table~\ref{tab:PGevolution}, for $N\geq2$, the point group of any ML-WTe$_2$ is $G_N=\mathrm{m}$ (a subgroup both of mm2 and of 2/m).
Inversion symmetry (and the horizontal $2_1$ screw axis) are lost for any ML-WTe$_2$ for the given stacking with a B-WTe$_2$ orthorhombic cell (they might be retrieved with a different stacking having an appropriate a monoclinic cell)}
\end{figure}

If $N<n_p$, this group-subgroup relation is, in general, not valid anymore: when looking at the point group, e.g., of a 1L, we have less conditions to satisfy (in particular, we remove the constraints on the specific stacking order of the layers).
Therefore, in general, the 1L point group could have more operations than the ML.
The example of ML graphite and graphene is discussed in the main text.
As another example, we consider WTe$_2$, as discussed in the caption of Fig.~\ref{fig:WTe2}.
\section{\label{app:C-LB-separation-monoclinic}An orthorhombic system where modes are not purely LB or C}
We now consider the system of Fig.~\ref{fig:alternation-examples}a.
2L-LM has point group 2/m (monoclinic, as any ML-LM with even $N$), while MLs with odd $N$ are orthorhombic.
By inspecting the crystal structure, we deduce that the unique axis of the 2L-LM is along $y$.
Therefore (see Table~\ref{tab:K-tensor-shape}), the force-constant tensor for 2L-LM has the form:
\begin{equation}
K^{(1)}_{\alpha\beta} = \begin{pmatrix}K_{11} & 0 & K_{31} \\ 0 & K_{22} & 0 \\ K_{31} & 0 & K_{33} \end{pmatrix}\label{eq:K1-monoclinic}
\end{equation}
for some non-zero values of the parameters $K_{11}$, $K_{22}$, $K_{33}$ and $K_{31}$.

In addition, the coincidence operation can be written, e.g., as a mirror orthogonal to $x$ followed by a translation along $z$, so that its rotational part is:
\begin{equation}
R=\begin{pmatrix}-1 & 0 & 0 \\ 0 & -1 & 0 \\ 0 & 0 & 1 \end{pmatrix}.
\end{equation}
This is not the only way to write the coincidence operation: composing it with any bulk operation still provides a valid coincidence operation.
The results discussed below, however, are independent of the specific choice.

$R$ and $K^{(1)}$ do not commute, therefore (see Sec.~\ref{sec:normal-modes}) force constants alternate at each interface, taking the values $K^{(1)}$ and $K^{(2)}$, the latter being defined as:
\begin{equation}
K^{(2)} = R K^{(1)} R^{-1} = \begin{pmatrix}K_{11} & 0 & -K_{31} \\ 0 & K_{22} & 0 \\ -K_{31} & 0 & K_{33} \end{pmatrix}.\label{eq:K2-monoclinic}
\end{equation}
We first observe that, if we limit to the $1\times 1$ block along $y$, we can consider $R$ and $K$ as commuting.
Therefore, there will be a mode with pure oscillations along $y$, i.e., a pure C mode.

Let us now focus only on the $xz$ subspace, and define the $xz$ sub-blocks of $K^{(1)}$ and $K^{(2)}$ as:
\begin{equation}
\hat K^{(1)}=\begin{pmatrix}K_{11} & K_{31}\\ K_{31} & K_{33} \end{pmatrix},\qquad
\hat K^{(2)}=\begin{pmatrix}K_{11} & -K_{31}\\ -K_{31} & K_{33} \end{pmatrix}.
\end{equation}
We first note that, in 2L-LM, the $x$ and $z$ component mix (due to the off-diagonal $K_{31}$ component), so that, as expected for a monoclinic system, we cannot define pure LB or C modes.
The same happens for all even $N$ (monoclinic).
One might expect that for odd $N$, since the point group is instead orthorhombic, $x$ (LB) and $z$ (C) modes would perfectly decouple.
However, this is not the case.
This can be verified by defining a displacement vector $\mathbf U = \left(u_x(1), u_z(1), u_x(2), u_z(2), \cdots, u_x(N), u_z(N)\right)^T$ so that the equation of motion Eq.~\eqref{eq:full1Dmodel} can be written as $-M\omega_n^2 \mathbf U = \hat K \mathbf U$, with $\hat K$ having the following block form:
\begin{equation}
\hat K = \begin{pmatrix}
        -\hat K^{(1)} & \hat K^{(1)} & 0 & \ldots & && 0\\
        \hat K^{(1)} & -\hat K^{(0)} & \hat K^{(2)} & 0 & \ldots && 0 \\
        0 & \hat K^{(2)} & -\hat K^{(0)} & \hat K^{(1)} & 0 & \ldots & 0 \\
        \vdots &&& \ddots &  \\
        0 & \ldots &&&& \hat K^{(i)} & -\hat K^{(i)}\\
\end{pmatrix}
\end{equation}
where $\hat K^{(0)} = \hat K^{(1)} + \hat K^{(2)}$ and $i=1$ for even $N$, while $i=2$ for odd $N$.

Even if the $\hat K^{(0)}$ block is diagonal, there are still mixed $xz$ components in the off-diagonal $\hat K^{(1)}$ and $\hat K^{(2)}$ blocks, and thus (independently of the parity of $N$) all eigenvectors have non-zero $x$ and $z$ components.
Nevertheless, for odd $N$ (orthorhombic) the matrix $\hat K$ commutes with the mirror operation orthogonal to $z$ at the centre of the ML, therefore eigenvectors can be chosen to be simultaneously eigenvectors also of this mirror operation (while this is not the case for even $N$).
Thus, the eigenvectors respect the orthorhombic symmetry of MLs with odd $N$, and the optical activity is given by the irreducible representations of the corresponding orthorhombic point group.

In summary, even in the orthorhombic case we cannot define purely LB and C modes (on the $xz$ subspace) and, more generally, the decoupling of the modes is determined by the crystal symmetry of 2L-LM, not of the full ML-LM.
The optical activity, on the other hand, is determined by the point group of the full ML-LM as discussed in the main text.

\section{\label{app:example-activity}Worked-out example: activity for group 2/m}
We now consider how to obtain the classification of the optical activity of the modes in the example of Fig.~\ref{fig:alternation-examples}a with $N=4$.
The symmetry of this system is described in Appendix~\ref{app:C-LB-separation-monoclinic} (ML-LM point group 2/m), and the four symmetry operations are: 1 (identity), 2 (180 degrees rotation about the $y$ axis), $\bar 1$ (inversion) and m (mirror plane, orthogonal to the $y$ axis).

There are two force-constant tensors, $K^{(1)}$ and $K^{(2)}$ in Eqs.~\eqref{eq:K1-monoclinic}, \eqref{eq:K2-monoclinic}, that alternate.
However, if we limit to the 1$\times$1 subspace for the decoupled C mode with oscillations along $y$, only a single component $K_{22}$ is sufficient to describe the force between any pair of layers.
We can therefore, for this specific case, use the model in Sec.~\ref{sec:normal-modes}, limiting to $\alpha=\beta=2$ ($y$ axis).
The final equation of motion can be written in matrix form as:
\begin{equation}
\omega_n^2 \mathbf U = \frac{k}{M}\left(\begin{matrix}
     1 & -1 &  0 &  0 \\
    -1 &  2 & -1 &  0 \\
     0 & -1 &  2 & -1 \\
     0 &  0 & -1 &  1 \\
\end{matrix}\right)\mathbf U\label{eq:example-equation-motion}
\end{equation}
where, as in the main text, we assume a harmonic form for $u(\ell,t)=u(\ell)e^{i\omega_n t}$, so that $\ddot u(\ell,t) = -\omega_n^2 u(\ell,t)$; $\mathbf U$ is the column vector of the displacements along $y$ for each layer, i.e., $\mathbf U = (u_y(\ell=1),u_y(\ell=2),u_y(\ell=3),u_y(\ell=4))^T$; and $k=K_{22}$.

Eq.~\eqref{eq:example-equation-motion} is a eigenvector equation with eigenvalues $\omega_n^2$, and can be solved to find the following 4 solutions ($\mathbf U^{(n)}$ being the corresponding eigenvectors):
\begin{equation}
\begin{cases}
    \omega_1^2 = 0, & \mathbf U^{(1)} = (\frac 1 2, \frac 1 2, \frac 1 2, \frac 1 2)^T \\
    \omega_2^2 = \left(2-\sqrt 2\right)\frac{k}{M}, & \mathbf U^{(2)} = (v', v'', -v'', -v')^T \\
    \omega_3^2 = 2\frac{k}{M}, & \mathbf U^{(3)} = (\frac 1 2, -\frac 1 2, -\frac 1 2, \frac 1 2)^T \\
    \omega_4^2 = \left(2+\sqrt 2\right)\frac{k}{M}, & \mathbf U^{(4)} = (-v'', v', -v', v'')^T
\end{cases},
\end{equation}
with $v'=\sqrt{\frac{2+\sqrt{2}}{8}}$ and $v''=\sqrt{\frac{2-\sqrt{2}}{8}}$.
The frequencies are the same as those obtained from Eq.~\eqref{eq:frequencies-chain}.

Now that we have the eigenvectors $\mathbf U^{(n)}$, in order to apply Eq.~\eqref{eq:optical-activity} we still need to get the table of the irreducible representations for the point group 2/m.
These are found, e.g., in Refs.~\onlinecite{Cracknell,Altmann1994}:
\begin{center}
\begin{tabular}{cccccc}
$\gamma$ & 1 & 2 & $\bar 1$ & m & functions \\  \hline
$A_g$ & $\phantom{-}1$ & $\phantom{-}1 $ & $\phantom{-}1$ & $\phantom{-}1$ & $x^2$, $y^2$, $z^2$, $xy$, $J_z$ \\
$B_g$ & $\phantom{-}1$ & $-1$ & $\phantom{-}1$ & $-1$ & $xz$, $yz$, $J_x$, $J_y$  \\
$A_u$ & $\phantom{-}1$ & $\phantom{-}1 $ & $-1$ & $-1$ & $z$  \\
$B_u$ & $\phantom{-}1$ & $-1$ & $-1$ & $\phantom{-}1$ & $x$, $y$
\end{tabular}
\end{center}
where the first row indicates the symmetry elements $g$, and the values in the table are the characters $\chi^{(\gamma)}(g)$ for the 4 irreducible representations $\gamma=A_g$, $B_g$, $A_u$ and $B_u$ of 2/m (they all have the same dimension $d_\gamma=1$, and the order of the 2/m group is $h=4$).
$A_g$ and $B_g$ are Raman active since they transform as quadratic functions, while $A_u$ and $B_u$ are IR active since they transform as linear functions.
Between the two Raman-active representations, only modes corresponding to $A_g$ are visible in a back-scattering geometry, because there are quadratic forms ($x^2$, $y^2$, $xy$) that involve only the $x$ and $y$ coordinates.

Applying Eq.~\eqref{eq:optical-activity} is straightforward, when we note that $\hat O_1$ is the identity; $\hat O_{2}$ is a 180 degrees rotation with axis along $y$, so it does not change the sign of displacements along $y$, but it swaps the order of the layers; $\hat O_{\bar 1}$ changes both the signs of the displacements and the order of the layers; and $\hat O_{m}$ changes the sign of displacements along $y$ but does not change the order of the layers, i.e.:
\begin{align}
    \hat O_{1}\left(
\right).
\end{align}
Applying Eq.~\eqref{eq:optical-activity}, we get that $p_\gamma(n=2)$ and $p_\gamma(n=4)$ are 1 only for $\gamma=B_g$ (i.e., Raman active only), while $p_\gamma(n=3)$ is 1 only for $\gamma=A_u$ (i.e., IR active only).
We skip $n=1$ as this is an acoustic mode with zero frequency where layers translate together of the same amount.
These representations correspond to the top row of the 2/m $(x)$ case of Table~\ref{tab:shear}.
\onecolumngrid
\section{Tables}
\begingroup
\LTcapwidth=\textwidth
\begin{longtable}{ l l c c c c c c c c}
\caption{
  Possible ML point groups $G_N$ that can be obtained knowing the space group and Hall number of the corresponding B-LM.
  Results for all settings compatible with a layered structure (e.g., discarding cubic space groups) and for different $n_{c}$ in the B-LM conventional cell, for systems of Category I.
  See Appendices~\ref{app:grouping-catI}, \ref{app:grouping-catIII} to apply the results of this table to Categories II, III.
  For each space group all inequivalent settings are considered and labeled by their Hall number.
  The bulk ($G_b$) and layer-invariant ($G_I$) point groups are also provided.
  $\diagup$ and $\times$ indicate that a LM with given Hall setting and $n_c$ cannot exist with our assumptions of being a MDO polytype (see Appendices~\ref{app:compatibility}, \ref{app:grouping-catI} for more details).
  Rhombohedral structures are only considered in their hexagonal setting.
  \label{tab:PGevolution}}\\
\toprule
Hall & Bulk space group & $G_b$ & $G_I$  & \multicolumn{6}{c}{Multilayer point group $G_N$} \\
number &  & &  & \multicolumn{6}{c}{$n_c$} \\
    & & & & 1 & 2 & 3 & 4 & 5 & 6 \\
\midrule
\endfirsthead
\multicolumn{10}{l}{\ldots continues from previous page.}\\
\toprule
Hall & Bulk space group & $G_b$ & $G_I$  & \multicolumn{6}{c}{Multilayer point group $G_N$} \\
number &  & &  & \multicolumn{6}{c}{$n_c$} \\
    & & & & 1 & 2 & 3 & 4 & 5 & 6 \\
\midrule
\endhead
\multicolumn{10}{r}{Continues in the next page\dots}\\
\endfoot
\bottomrule
\endlastfoot
&\multicolumn{9}{l}{Triclinic}\\
\midrule
1 & 1 (P$1$) & $1$ & $1$ & $1$ & $1$ & $1$ & $1$ & $1$ & $1$ \\
\noalign{\global\arrayrulewidth=0.5px}\arrayrulecolor{gray}\midrule\arrayrulecolor{black}\noalign{\global\arrayrulewidth=1px}
2 & 2 (P$\bar1$) & $\bar1$ & $1$ & $\bar1$ & $1$ or $\bar1$ & $\times$ & $\times$ & $\times$ & $\times$ \\
\midrule
&\multicolumn{9}{l}{Monoclinic}\\
\midrule
3 & 3 (P$121$) & $2$ & $1$ & $2$ & $1$ or $2$ & $\times$ & $\times$ & $\times$ & $\times$ \\
4 & 3 (P$112$) & $2$ & $2$ & $2$ & $2$ & $2$ & $2$ & $2$ & $2$ \\
&\color{gray}\itshape Equivalent Hall numbers: & \multicolumn{8}{l}{\color{gray}5 (P$211$) $\rightarrow$ 3}\\
\noalign{\global\arrayrulewidth=0.5px}\arrayrulecolor{gray}\midrule\arrayrulecolor{black}\noalign{\global\arrayrulewidth=1px}
6 & 4 (P$12_11$) & $2$ & $1$ & $2$ & $1$ or $2$ & $\times$ & $\times$ & $\times$ & $\times$ \\
\end{longtable}

\endgroup 
\begingroup
\LTcapwidth=0.9\textwidth
\begin{longtable}{|c|c|cc|ccc|cccc|ccccc|cccccc|}
\caption{
  LBM classification for a $N$-layer ML to its point group $G_N$.
  For a given point group and $N$, the modes are reported from left to right in order of increasing frequency.
  Raman active modes are denoted as $\ocircle$, IR active ones as $\times$, while those that are both Raman and IR active as $\otimes$.
  A red symbol indicates that the mode can be detected in a backscattering Raman experiment orthogonal to the layers.
  The irreducible representation to which each mode belongs is also reported.
  Whenever necessary, different orientations of the principal symmetry element with respect to $z$ are considered and specified in parenthesis.
  In cases where it is not possible to decouple C and LB modes, we report them noting that the mode assignment coincides with that of the corresponding mixed mode in Table\ref{tab:shear}.
}\\
\toprule
Pointgroup &\multicolumn{1}{|c|}{$N=2$} & \multicolumn{2}{|c|}{$N=3$} & \multicolumn{3}{|c|}{$N=4$} & \multicolumn{4}{|c|}{$N=5$} & \multicolumn{5}{|c|}{$N=6$} & \multicolumn{6}{|c|}{$N=7$}\\
\midrule
\endfirsthead
\multicolumn{22}{l}{\dots continues from previous page.}\\
\toprule
Pointgroup &\multicolumn{1}{|c|}{$N=2$} & \multicolumn{2}{|c|}{$N=3$} & \multicolumn{3}{|c|}{$N=4$} & \multicolumn{4}{|c|}{$N=5$} & \multicolumn{5}{|c|}{$N=6$} & \multicolumn{6}{|c|}{$N=7$}\\
\midrule
\endhead
\multicolumn{22}{r}{Continues in the next page\dots}\\
\endfoot
\bottomrule
\endlastfoot
 \multirow{2}{*}{$1$ } & \color{red}{$\bm\otimes$}& \color{red}{$\bm\otimes$}& \color{red}{$\bm\otimes$}& \color{red}{$\bm\otimes$}& \color{red}{$\bm\otimes$}& \color{red}{$\bm\otimes$}& \color{red}{$\bm\otimes$}& \color{red}{$\bm\otimes$}& \color{red}{$\bm\otimes$}& \color{red}{$\bm\otimes$}& \color{red}{$\bm\otimes$}& \color{red}{$\bm\otimes$}& \color{red}{$\bm\otimes$}& \color{red}{$\bm\otimes$}& \color{red}{$\bm\otimes$}& \color{red}{$\bm\otimes$}& \color{red}{$\bm\otimes$}& \color{red}{$\bm\otimes$}& \color{red}{$\bm\otimes$}& \color{red}{$\bm\otimes$}& \color{red}{$\bm\otimes$}\\*
  & {\footnotesize  $A_{}$}& {\footnotesize  $A_{}$}& {\footnotesize  $A_{}$}& {\footnotesize  $A_{}$}& {\footnotesize  $A_{}$}& {\footnotesize  $A_{}$}& {\footnotesize  $A_{}$}& {\footnotesize  $A_{}$}& {\footnotesize  $A_{}$}& {\footnotesize  $A_{}$}& {\footnotesize  $A_{}$}& {\footnotesize  $A_{}$}& {\footnotesize  $A_{}$}& {\footnotesize  $A_{}$}& {\footnotesize  $A_{}$}& {\footnotesize  $A_{}$}& {\footnotesize  $A_{}$}& {\footnotesize  $A_{}$}& {\footnotesize  $A_{}$}& {\footnotesize  $A_{}$}& {\footnotesize  $A_{}$}\\
\midrule
 \multirow{2}{*}{$\bar1$ } & \color{red}{$\bm\ocircle$}& \color{red}{$\bm\ocircle$}& $\times$& \color{red}{$\bm\ocircle$}& $\times$& \color{red}{$\bm\ocircle$}& \color{red}{$\bm\ocircle$}& $\times$& \color{red}{$\bm\ocircle$}& $\times$& \color{red}{$\bm\ocircle$}& $\times$& \color{red}{$\bm\ocircle$}& $\times$& \color{red}{$\bm\ocircle$}& \color{red}{$\bm\ocircle$}& $\times$& \color{red}{$\bm\ocircle$}& $\times$& \color{red}{$\bm\ocircle$}& $\times$\\
\end{longtable}
\endgroup

\begingroup
\LTcapwidth=0.9\textwidth
\begin{longtable}{|c|c|cc|ccc|cccc|ccccc|cccccc|}
  \caption{
  C modes classification for a $N$-layer ML according to its point group $G_N$.
  For a given point group and $N$, the modes are reported from left to right in order of increasing frequency.
  Raman active modes are denoted as $\ocircle$, IR active ones as $\times$, while those that are both Raman and IR active as $\otimes$.
  A red symbol indicates that the Raman mode can be detected in a backscattering Raman experiment orthogonal to the layers.
  The irreducible representation to which each mode belongs is also reported.
  Whenever necessary, different orientations of the principal symmetry element with respect to the layering direction ($z$) are considered and specified in parenthesis.
  In cases where it is not possible to decouple C and LB modes, we still report them, and we note that the mode assignment coincides with that of the corresponding mixed mode in Table~\ref{tab:layerbreathing}.
  Only modes along the first principal direction are reported if the pattern of Raman/IR activity is the same as for modes along second direction, and the irreducible representations differ just by a naming convention (e.g., $B_2$ vs $B_3$), otherwise displacements in both principal directions are shown.
}\\
\toprule
Pointgroup &\multicolumn{1}{|c|}{$N=2$} & \multicolumn{2}{|c|}{$N=3$} & \multicolumn{3}{|c|}{$N=4$} & \multicolumn{4}{|c|}{$N=5$} & \multicolumn{5}{|c|}{$N=6$} & \multicolumn{6}{|c|}{$N=7$}\\
\midrule
\endfirsthead
\multicolumn{22}{l}{\dots continues from previous page.}\\
\toprule
Pointgroup &\multicolumn{1}{|c|}{$N=2$} & \multicolumn{2}{|c|}{$N=3$} & \multicolumn{3}{|c|}{$N=4$} & \multicolumn{4}{|c|}{$N=5$} & \multicolumn{5}{|c|}{$N=6$} & \multicolumn{6}{|c|}{$N=7$}\\
\midrule
\endhead
\multicolumn{22}{r}{Continues in the next page\dots}\\
\endfoot
\bottomrule
\endlastfoot
 \multirow{2}{*}{$1$ } & \color{red}{$\otimes$}& \color{red}{$\otimes$}& \color{red}{$\otimes$}& \color{red}{$\otimes$}& \color{red}{$\otimes$}& \color{red}{$\otimes$}& \color{red}{$\otimes$}& \color{red}{$\otimes$}& \color{red}{$\otimes$}& \color{red}{$\otimes$}& \color{red}{$\otimes$}& \color{red}{$\otimes$}& \color{red}{$\otimes$}& \color{red}{$\otimes$}& \color{red}{$\otimes$}& \color{red}{$\otimes$}& \color{red}{$\otimes$}& \color{red}{$\otimes$}& \color{red}{$\otimes$}& \color{red}{$\otimes$}& \color{red}{$\otimes$}\\*
  & {\footnotesize  $A_{}$}& {\footnotesize  $A_{}$}& {\footnotesize  $A_{}$}& {\footnotesize  $A_{}$}& {\footnotesize  $A_{}$}& {\footnotesize  $A_{}$}& {\footnotesize  $A_{}$}& {\footnotesize  $A_{}$}& {\footnotesize  $A_{}$}& {\footnotesize  $A_{}$}& {\footnotesize  $A_{}$}& {\footnotesize  $A_{}$}& {\footnotesize  $A_{}$}& {\footnotesize  $A_{}$}& {\footnotesize  $A_{}$}& {\footnotesize  $A_{}$}& {\footnotesize  $A_{}$}& {\footnotesize  $A_{}$}& {\footnotesize  $A_{}$}& {\footnotesize  $A_{}$}& {\footnotesize  $A_{}$}\\*
\midrule
 \multirow{2}{*}{$\bar1$ } & \color{red}{$\ocircle$}& \color{red}{$\ocircle$}& $\times$& \color{red}{$\ocircle$}& $\times$& \color{red}{$\ocircle$}& \color{red}{$\ocircle$}& $\times$& \color{red}{$\ocircle$}& $\times$& \color{red}{$\ocircle$}& $\times$& \color{red}{$\ocircle$}& $\times$& \color{red}{$\ocircle$}& \color{red}{$\ocircle$}& $\times$& \color{red}{$\ocircle$}& $\times$& \color{red}{$\ocircle$}& $\times$\\
\end{longtable}
\endgroup

\twocolumngrid

%

\end{document}